\documentclass[aps,floatfix,prd,nofootinbib,superscriptaddress,reprint,showpacs
  ,10pt,preprintnumbers,longbibliography]{revtex4-1}
\usepackage[utf8]{inputenc}
\usepackage[pdftex]{graphicx}
\usepackage{float}
\usepackage{amsmath}
\usepackage{amssymb}
\usepackage{mathtools}
\usepackage{braket}
\usepackage{siunitx}
\usepackage{amsfonts,amsthm}
\usepackage{dsfont}
\usepackage{array}
\usepackage{bm}
\usepackage{mathrsfs}
\usepackage{pifont}
\usepackage{multirow}
\usepackage{upgreek}
\usepackage[dvipsnames]{xcolor}
\usepackage[pdftex,
  pdftitle={},
  pdfauthor={},
  bookmarks,
  colorlinks,
  linkcolor=myblue,
  citecolor=mymagenta,
  menucolor=black,
  urlcolor=myblue,
  plainpages=false,
  pdfpagelabels,
  hypertexnames=false]{hyperref}
\usepackage{verbatim}
\usepackage{slashed}
\usepackage{cleveref}
\usepackage{ucs}
\usepackage{subcaption}
\usepackage{csquotes}
\usepackage{bbold}
\usepackage{tikz}
\usetikzlibrary{shapes.geometric, shapes.misc}

\makeatletter
\let\newfloat\newfloat@ltx
\makeatother
\usepackage{algorithm}
\makeatletter
\renewcommand{\ALG@name}{Algorithm }
\makeatother
\usepackage[noend]{algpseudocode}

\usepackage[normalem]{ulem}

\DeclareMathOperator{\im}{i}

\newcommand{\md}{\ensuremath{\mathrm{d}}}

\newcommand{\br}[1]{\ensuremath{\left(#1\right)}}

\newcommand{\brc}[1]{\ensuremath{\left\{#1\right\}}}

\theoremstyle{definition}
\newtheorem{definition}{Definition}[section]

\newtheorem{lemma}[definition]{Lemma}

\newtheorem{theorem}{Theorem}
\newtheorem{corollary}[definition]{Corollary}

\theoremstyle{remark}

\definecolor{mymagenta}{RGB}{200, 0, 100}
\definecolor{myblue}{RGB}{45, 48, 146}

\graphicspath{{plots/}}

\begin{document}

\title{No-go theorems for dual Hamiltonians of Non-Abelian Lattice Gauge Theories}

\author{Marco Garofalo}
\affiliation{Helmholtz-Institut f\"ur Strahlen- und Kernphysik, University of
  Bonn, Nussallee 14-16, 53115 Bonn, Germany}
\affiliation{Bethe Center for Theoretical Physics, University of Bonn,
  Nussallee 12, 53115 Bonn, Germany}

\author{Tobias Hartung}
\affiliation{Northeastern University - London, Devon House, St Katharine Docks,
  London, E1W 1LP, United Kingdom}
\affiliation{Khoury College of Computer Sciences, Northeastern University,
  \#202, West Village Residence Complex H, 440 Huntington Ave, Boston, MA
  02115,
  USA}

\author{Timo Jakobs}
\affiliation{Helmholtz-Institut f\"ur Strahlen- und Kernphysik, University of
  Bonn, Nussallee 14-16, 53115 Bonn, Germany}
\affiliation{Bethe Center for Theoretical Physics, University of Bonn,
  Nussallee 12, 53115 Bonn, Germany}

\author{Paul Ludwig}
\affiliation{Helmholtz-Institut f\"ur Strahlen- und Kernphysik, University of
  Bonn, Nussallee 14-16, 53115 Bonn, Germany}
\affiliation{Bethe Center for Theoretical Physics, University of Bonn,
  Nussallee 12, 53115 Bonn, Germany}

\author{Johann Ostmeyer}
\affiliation{Helmholtz-Institut f\"ur Strahlen- und Kernphysik, University of
  Bonn, Nussallee 14-16, 53115 Bonn, Germany}
\affiliation{Bethe Center for Theoretical Physics, University of Bonn,
  Nussallee 12, 53115 Bonn, Germany}

\author{Carsten Urbach}
\affiliation{Helmholtz-Institut f\"ur Strahlen- und Kernphysik, University of
  Bonn, Nussallee 14-16, 53115 Bonn, Germany}
\affiliation{Bethe Center for Theoretical Physics, University of Bonn,
  Nussallee 12, 53115 Bonn, Germany}
\date{\today}

\begin{abstract}
  Hamiltonian simulations of lattice gauge theories promise new insights
  into the inner workings of QCD. The rapid development of quantum computing
  hardware suggests that large scale simulations may soon become feasible.
  However, no
  efficient formulation suitable for simulations near the continuum
  limit is currently known for
  the non-Abelian case, and existing candidates typically feature long
  range interactions in the electric part of the
  Hamiltonian.
  We prove that such non-localities are unavoidable when the gauge
  configuration is reparametrised while preserving both the number of
  degrees of freedom and the fundamental commutation relations.  
  We then attempt to mitigate this by introducing
  additional gauge links alongside additional constraints. Although
  this approach is successful in U(1), we prove that the corresponding
  construction is inconsistent for non-Abelian lattice gauge theories.
\end{abstract}

\maketitle

\section{Introduction}

The Hamiltonian operator $\hat H$ plays a pivotal role in any quantum 
(field) theory and, thus, for our description of fundamental
interactions in nature: $\hat H$ entails the symmetries of the
physical system, determines the system's energy eigenstates, and 
governs time evolution via the time evolution operator.
For any quantum field theory it is, therefore, of utmost importance to
control $\hat H$ as well as possible, analytically or
numerically. Optimally, one is able to perform a change of variables
such that observables of interest can be computed analytically, or at
least much more easily than in the original formulation.

Gauge field theories represent one family of quantum field theories
of particular interest, because gauge symmetries lie at the core of
the three fundamental interactions the Standard Model (SM) of particle
physics is composed of. Many such gauge theories are inherently
non-perturbative, a property they share with the strong interaction in
the SM, which comprises a local SU$(3)$ gauge symmetry. As
a consequence, a non-perturbative treatment of the corresponding
Hamiltonian is required. 

Since quantum field theories in general also require renormalisation,
the non-perturbative method one resorts to is the lattice gauge theory
approach. The corresponding lattice Hamiltonian was found by Kogut and
Susskind in 1975~\cite{Kogut:1974ag}, and it is special in at least
two regards: first, it 
contains gauge field operators connecting sites of a discrete spatial
lattice. These are combined and traced over as smallest loops --
so-called plaquettes on this very lattice in the potential (or
so-called magnetic) part of the Hamiltonian. Second, $\hat H$ is
subject to additional constraints which ensure (colour) charge
conservation. These Gauss' law constraints on link and canonical momentum
operators exist for each lattice site and reduce the total Hilbert
space to the physical Hilbert space. 

It turns out that it would be advantageous for many applications and
in particular for the continuum limit to work in a
basis specifically tailored for the magnetic part of the
Hamiltonian. Such approaches work superbly in Abelian U$(1)$ lattice
gauge theories~\cite{Miranda-Riaza2026,Haase:2020kaj,Jakobs:2025U1}: a
change of variables to a rotor basis can be performed which 
leads to a free magnetic part involving exclusively terms with only
operators at lattice site $\mathbf{x}$, and an electric part
containing next-neighbour interactions, by implementing Gauss' law as
static constraints to reduce the number of degrees of freedom
(dofs) to a minimum.

Despite significant effort~\cite{Raychowdhury:2019iki,PhysRevD.107.074504,Jakobs:2023lpp,Bauer:2023jvw,Grabowska:2024emw,Burbano:2024uvn,Jakobs:2025rvz,Kane:2025ybw,Ciavarella:2025bsg,Froland:2026aff,Gupta:2026tcg},
the equivalent construction is not known
for non-Abelian lattice gauge theories. In fact, we will prove a no-go 
theorem below that is at least part of the reason for this. Simply put, 
for non-Abelian gauge theories it is not possible to canonically 
transform the original Kogut-Susskind Hamiltonian into a new one (i.e.\@ 
while preserving the number of degrees-of-freedom and the fundamental 
commutators) which has both 
\begin{enumerate}
\item a free magnetic part,
\item only local interactions: only terms involving products of (non-trivial) operators at
  lattice sites $\mathbf{x}$ and $\mathbf{y}$ with maximal distance
  $|\mathbf{x}-\mathbf{y}|$ bounded by a constant independent of the
  lattice extent $\ell$.
\end{enumerate}

In addition, we show that a class of what we call dynamic constraints
-- which leads in U(1) to the same rotor formulation as obtained with
static constraints, is fundamentally inconsistent in non-Abelian gauge
theories.

\section{Theory}
\label{sec:theory}

The subject of interest in this article is the Kogut-Susskind Hamiltonian,
first proposed in \cite{Kogut:1974ag}. In the following we will recall the
operators required to define it and outline its main features. For more details
we recommend the theory section of \cite{Chin1985} or our previous publications
on the topic \cite{Jakobs:2025rvz,Jakobs:2025U1}.

The Kogut-Susskind Hamiltonian is defined on a cubical spatial lattice
$\Lambda = \{1,\ldots,\ell\}^d$ in $d$ dimensions with extent $\ell$.
In the following we will assume $d>1$.
Contrary to its Lagrangian counterpart time is not discretised. For each
lattice site there are $d$ gauge fields, where $d$ is the number of spatial
dimensions of the theory. Thus, the gauge fields will be indexed by a position
$\mathbf{x}$ and a direction $i$. Classically, the gauge field at
$\hat{U}_{\mathbf{x}, i}$ can be understood as the local gauge transformation
connecting the sites $\mathbf{x}$ and $\mathbf{x} + a \hat{\mathbf{i}}$, where
$a$ denotes the lattice spacing and $\hat{\mathbf{i}}$ the unit vector
in direction $i$.

In the following we will treat the Kogut-Susskind theory through the lens of a
quantum many-body problem. That is, every gauge link $U_{\mathbf{x}, i}$ in
$\Lambda$ is viewed as a quantum particle on the group manifold of the gauge
group $G$. Quantum states of the theory are described by a wave function
\begin{equation}
    \psi\left(\dots, {U_{\mathbf{x}, k}}, \dots \right): G^{N_{\textrm{links}}}
    \, \,
    \rightarrow
    \, \, \mathbb{C} \, ,
\end{equation}
assigning a complex probability amplitude to every classical configuration of
gauge links $\left\{U_{\mathbf{x}, i}\right\}$.
To formulate a Hamiltonian for this many-body system we begin by introducing
the left and right momentum operators $\hat{L}^c_{\mathbf{x},k}$ and
$\hat{R}^c_{\mathbf{x},k}$, respectively, which take the shape of
Lie derivatives:
\begin{align}
    \hat{L}^c_{\mathbf{x},k} \, \psi \	 & = -\im
    \frac{\mathrm{d}}{\mathrm{d}\alpha}\,
    \psi \left(\dots, e^{- \im \alpha \tau_c}\, U_{\mathbf{x},k},
    \dots\right)|_{\alpha = 0}\,
    \intertext{and}
    \hat{R}^c_{\mathbf{x},k} \, \psi \	 & = -\im
    \frac{\mathrm{d}}{\mathrm{d}\alpha}\,
    \psi \left(\dots,U_{\mathbf{x},k} \, e^{\im \alpha
        \tau_c},\dots\right)|_{\alpha = 0}\, .
\end{align}
Here, the $\tau_c$ denote the generators of the Lie algebra in colour space.
With this we can define the kinetic term (also commonly referred to as the
\emph{electric} term) of the theory. It is implemented by summing the squared
momentum operators of all particles:
\begin{equation}
    \hat{H}_{\textrm{kin}} = \frac{g^2}{2}\sum_{\mathbf{x},c,k}
    \left(\hat{L}_{\mathbf{x},k}^c\right)^2 \, ,\label{eq:kinetic}
\end{equation}
where $g^2$ is the bare gauge coupling of the theory. To introduce
non-trivial dynamics, a potential term needs to be added to the Hamiltonian. For this
one introduces the position or link operators $\hat{U}_{\mathbf{x}, i}$, which
are defined as
\begin{equation}
    \hat{U}_{\mathbf{x}, i} \, \psi = U_{\mathbf{x}, i}\
    \psi\left(\dots, {U_{\mathbf{x}, i}}, \dots\right) \, .
\end{equation}
As an interacting potential one chooses, similar to the Lagrangian counterpart,
the trace of a fundamental oriented loop of gauge links called the plaquette.
It is defined as
\begin{equation}
    \label{eq:plaquette}
    \hat{P}_{\mathbf{x}, i j} =\ \hat{U}^{~}_{\mathbf{x},i}\,
    \hat{U}^{~}_{\mathbf{x}+\hat{\mathbf{i}}, j}\,
    \hat{U}^\dagger_{\mathbf{x}+\hat{\mathbf{j}}, i}\,
    \hat{U}^\dagger_{\mathbf{x},j} \, .
\end{equation}
With this the full Hamiltonian reads
\begin{equation}
    \label{eq:hamiltonian}
    \begin{split}
        \hat H & =\ \frac{g^2}{2}\sum_{\mathbf{x},c,k}
        \left(\hat{L}_{\mathbf{x},k}^c\right)^2 +
        \frac{2}{g^2}\sum_{\mathbf{x},j<i}
        \mathrm{Tr} \left[\,
        \mathbb{1} -
        \mathrm{Re}\,
        \hat{P}_{\mathbf{x}, ij} \right] \, .
    \end{split}
\end{equation}
The second term is also commonly referred to as the \emph{magnetic} term.

It is easy to show that the Hamiltonian commutes with the Gauss law operators
\begin{equation}
    \hat{G}^c_{\mathbf{x}} = \sum_{i} \left(
    \hat{L}^c_{\mathbf{x},i} + \hat{R}^c_{\mathbf{x} -
        \hat{\mathbf{i}},i} \right) \, .
\end{equation}
They are the generators of local gauge transformations.
Physical states are required to be gauge invariant. I.e.\ they fulfil
at each lattice site $\mathbf{x}$
\begin{equation}
    \hat{G}^c_{\mathbf{x}} \ket{\psi} = \sum_{k} \left(
    \hat{L}^c_{\mathbf{x},k} + \hat{R}^c_{\mathbf{x} -
        \hat{\mathbf{k}},k} \right) \ket{\psi} = Q_\mathbf{x}\ket{\psi}\,,
\end{equation}
with $Q_\mathbf{x}$ the charge, which we set to zero in the following.
Lastly it is useful to recapitulate the commutation relations of the
operators used. Similarly to regular quantum mechanics, the position and
momentum operators of the theory do not commute. They obey
\begin{align}
    [\hat{L}^c_{\mathbf{x},i}, \hat{U}_{\mathbf{y},j}]\  & =\
    -\delta_{\mathbf{x}
    \mathbf{y}} \, \delta_{ij} \, \tau_c\, \hat{U}_{\mathbf{x},i}\,, \\
    [\hat{R}^c_{\mathbf{x},i}, \hat{U}_{\mathbf{y},j}]\  & =\
    \delta_{\mathbf{x}
        \mathbf{y}} \, \delta_{ij} \, \hat{U}_{\mathbf{x},i} \, \tau_c \, .
\end{align}
For non-Abelian gauge groups the momentum operators also pick up non-trivial
commutation relations with each other. They read
\begin{align}\label{eq:L}
    [\hat{L}^a_{\mathbf{x},i}, \hat{L}^b_{\mathbf{y},j}] & = \im f_{abc}
    \, \delta_{\mathbf{x}
    \mathbf{y}} \, \delta_{ij} \, \hat{L}^c_{\mathbf{x},i} \,   ,                       \\
    \label{eq:R}
    [\hat{R}^a_{\mathbf{x},i}, \hat{R}^b_{\mathbf{y},j}] & = \im f_{abc}
    \, \delta_{\mathbf{x}
        \mathbf{y}} \, \delta_{ij} \, \hat{R}^c_{\mathbf{x}.i} \, ,
\end{align}
where $f_{abc}$ are the structure constants of the gauge group. Left and right
momentum operators do commute.

\subsection{Locality and related conventions}

Since locality and derived properties are central to this work, we will now
introduce our nomenclature. Specifically, we distinguish the following terms:
\begin{itemize}
    \item \textit{Free}: single-site terms like eq.~\eqref{eq:kinetic} (technically 0-local).
    \item \textit{Local}: any term that involves connections with bounded distances, more specifically \textit{$k$-local} if $k$ is the maximal possible distance (e.g.\ the nearest neighbour term in eq.~\eqref{eq:mixing-rotor} is 1-local).
    \item \textit{Non-local}: connection distances within a term grow with the lattice size $\ell$ and diverge as $\ell\rightarrow\infty$. We focus on \textit{polynomially non-local} terms where distances grow as $\ell^\alpha$ with $\alpha>0$.
    \item \textit{$n$-point} terms: counts number of operators that are multiplied (e.g.\ eq.~\eqref{eq:hamiltonian} is 2-point + 4-point).
    \item \textit{Many-point} terms: $n$-point terms where $n$ grows with lattice size without bound.
\end{itemize}
Note that we do not consider interaction strength in this convention,
as the strength (unless it is zero) is not relevant in this work, but
only the distance.

\subsection{Dual Formulation}

To obtain continuum physics the weak coupling limit $g^2 \rightarrow 0$ of the
theory must be taken. In this limit the 4-point interaction of the magnetic
term dictates the behaviour of the theory. Here, the wave function of a low
energy state will only be non-vanishing for gauge configurations where $P
    \approx \mathbb{1}$ holds for all plaquettes. This transition is expected to
get sharper with weaker couplings.

A common strategy to digitise the Hilbert space of the theory is to expand the
wave function in the low energy eigenstates of the momentum operators
\cite{PhysRevD.91.054506}. At weak couplings, these Fourier like expansions
require an ever-growing number of basis states to accurately reproduce the
sharp transitions within the ground state. Thus, a more suitable set of basis
functions is needed. This, however, is difficult to achieve in the original link
formulation of the theory, as the magnetic term does not meaningfully constrain
the occupied states of any individual gauge link.

Therefore, one aims to reparametrise the theory in terms of plaquette degrees
of freedom.
In such \textit{dual} reparametrisations the magnetic term becomes free and
efficient sets of basis functions are well known. E.g.\ the ground state of the
pure gauge theory is well approximated by the product state
\begin{equation}
    \psi_0 \approx \prod_{P \in \Lambda} \phi_0 (P)
\end{equation}
where $\phi_0$ is the ground state of a single plaquette system with open
boundary conditions. Tests at small system sizes further suggest that
the Fock space spanned by the low energy states of the single plaquette
approximate the Hilbert space of the many-plaquette theory efficiently
\cite{Jakobs:2025U1,Miranda-Riaza2026,Fontana:2024rux}.

Here the error introduced by truncating the Hilbert space is largely
independent of the coupling and no strong increase in basis size is observed.
Therefore, the authors believe that reformulating the Kogut-Susskind
Hamiltonian in terms of plaquette degrees of freedom would be the way
forward when it comes to Hamiltonian simulations of lattice gauge theories at weak
couplings.

\subsection{Enlarged Hilbert space and constraints}
\label{sec:hilbert-space-constraints}

In the following we will explore possible routes to such a
magnetically free formulation. In order for a theory to be magnetically free it
must contain rotor degrees of freedom $\hat W_{\mathbf{x}, ij}$ such that
\begin{equation}
    \hat{H}_{\textrm{magnetic}} = \frac{2}{g^2}\sum_{\mathbf{x},j<i}
    \mathrm{Tr} \left[\,
    \mathbb{1} -
    \mathrm{Re}\,
    \hat{W}_{\mathbf{x}, ij} \right] \, .
\end{equation}
Clearly, the number of rotors $N_\text{rotors}\le N_{\textrm{plaquettes}}$ is no larger than the number of plaquettes.
The rotors $\hat{W}_{\mathbf{x}, ij}$ have associated canonical momentum operators
$\hat{L}^c_{\mathbf{x}, ij}$ and $\hat{R}^c_{\mathbf{x}, ij}$. They fulfil the same
canonical commutation relations as a gauge link.

For the remaining degrees of freedom we get to choose $N_{\textrm{links}} -
    N_{\textrm{rotors}}$ of the original gauge links. In the following we will
refer to these as helper links. In order for all the original links to be
captured by the rotors and helper links, the helper links need to form a
maximal tree connecting all sites of the lattice (plus a single additional helper link generating a loop for periodic boundary conditions).
The number of rotors $N_\text{rotors}$ required to fulfil this condition is quantified in \cref{th:max-tree}.
In 2D we find exactly $N_\text{rotors}= N_{\textrm{plaquettes}}$.
We will refer to such a
choice of helper links as a \textit{static constraint}.

In this paper we will also consider another approach discussed more in
\cref{sec:dynamic-constraints}. Here we set out to keep all helper
links active after
the introduction of the rotors. As we now have too many degrees of freedom, we
introduce $N_{\textrm{plaquettes}}$ additional local plaquette constraints.
These will be referred to as \textit{dynamic constraints}, as the helper links retain non-trivial dynamics.

\subsection{Example: the Rotor Hamiltonian in 2D
 with static constraints}

\begin{figure}
    \centering
    \begin{subfigure}[c]{.47\columnwidth}
        \includegraphics[width=\textwidth]{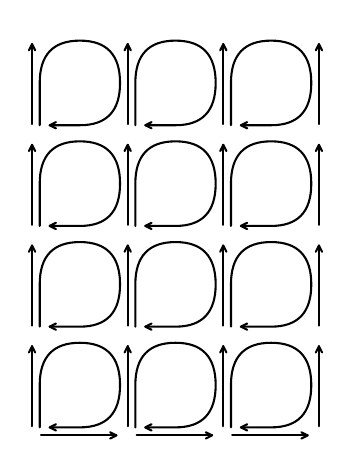}
        \caption{The choice of static constraint found in \cite{PhysRevD.107.074504}.}
        \label{fig:mathur-max-tree-lattice}
    \end{subfigure}
    \hfill
    \begin{subfigure}[c]{.47\columnwidth}
        \includegraphics[width=\textwidth]{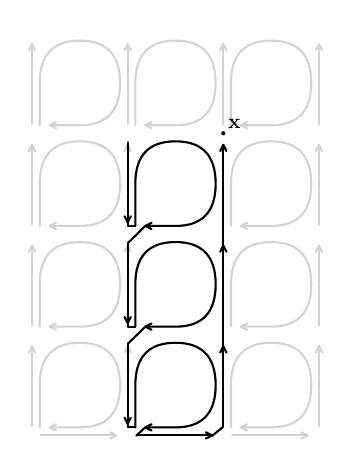}
        \caption{The link products $\mathcal{S}_{\mathbf{x}}$ replacing $U_{\mathbf{x},1}$.}
        \label{fig:mathur-max-tree-path}
    \end{subfigure}
    \caption{Visualisation of the new rotor degrees of freedom in 2D with open boundary conditions.
    	After applying static constraints, the remaining helper links form a maximal tree.}
\end{figure}

Let us elaborate on static constraints using an example by considering
a pure SU(N) gauge theory in two spatial dimensions with open boundary
conditions.
A possible choice of static constraints in open boundary conditions can be
found in Ref.~\cite{PhysRevD.107.074504}: the gauge degrees of freedom
remaining with that choice are depicted in
\cref{fig:mathur-max-tree-lattice} and the
reparametrised Hamiltonian reads
\begin{equation}
    \begin{split}
        \hat{H} = \frac{g^2}{2} & \sum_{\mathbf{x},c} \left[ \left(
            \hat{R}_{\mathbf{x}}^c +
            \mathcal{R}(U_{\mathbf{x},2}^\dagger)^c_{\,c'}
            \hat{L}^{c'}_{\mathbf{x} - \hat{\mathbf{2}}} +
        \delta_{x_2,1} \hat{L}^c_{\mathbf{x},1} \right)^2 \right.                                                 \\
                                & \qquad \quad \left. + \left( \hat{L}^c_{\mathbf{x}, 2} + \hat{L}^c_{\mathbf{x}}
            + \mathcal{R}(\mathcal{S}_\mathbf{x}^\dagger)^c_{\,c'}
            \hat{R}^{c'}_{\mathbf{x} - \hat{\mathbf{1}}}
        \right)^2 \right]                                                                                         \\
                                & \qquad \qquad + \frac{2}{g^2}\sum_{\mathbf{x}}
        \mathrm{Tr} \left[\, \mathbb{1} -
        \mathrm{Re}\, \hat{W}_{\mathbf{x}} \right] \, .
    \end{split}
\end{equation}
Here, $\mathcal{S}_{\mathbf{x}}$ denotes the product of gauge degrees
of freedom along a path connecting $\mathbf{x}$ and $\mathbf{x} -
\hat{\mathbf{1}}$ as depicted in \cref{fig:mathur-max-tree-path}.
This path is required, because the interaction mediated by
the original link $U_{\mathbf{x},1}$ needs to be now mediated by the
remaining gauge degrees of freedom. Also used is the colour rotation
matrix 
\begin{equation}
    \mathcal{R}^c_{\,c'}(U) = 2 \mathrm{Tr} \left[ \tau_c U \tau_{c'}
        U^\dag \right] \, .
\end{equation}

Gauss' law now takes the form
\begin{equation}
    \left( \hat{L}_{\mathbf{x}, 2} + \hat{R}_{\mathbf{x} - \hat{\mathbf{2}}, 2} + \hat{L}^c_{\mathbf{x}, ij} + \hat{R}^c_{\mathbf{x}, ij} \right) \ket{\psi} = 0
\end{equation}
for sites where $x_2 > 1$ and
\begin{equation}
    \left( \hat{L}_{\mathbf{x}, 2} + \hat{L}_{\mathbf{x}, 1} + \hat{R}_{\mathbf{x} - \hat{\mathbf{1}}, 1} + \hat{L}^c_{\mathbf{x}} + \hat{R}^c_{\mathbf{x}} \right) \ket{\psi} = 0
\end{equation}
for sites at the bottom of the lattice where $x_2=1$.

\subsubsection{Simplifications in U(1)}

As in U(1) $\hat{L} = -\hat{R}$ holds, the rotor degrees of freedom
cancel out in Gauss' law. For vertices at the top of the lattice
Gauss' law then reads 
\begin{equation}
    \hat{L}_{\mathbf{x}, 2} \ket{\psi_{\textrm{phys}}} = 0 \, .
\end{equation}
forcing the links at the top into the electric ground state. The same
logic can now be applied to the helper links one row below the top. In
this way one can iteratively remove all helper links from the
system. Left is the well known rotor formulation of U(1) (se also
Ref.~\cite{Kaplan:2018vnj}):
\begin{equation}
    \hat{H}_{\textrm{rotor}} = \frac{g^2}{2} \sum_{\mathbf{x}, i} \left(
    \hat{L}_{\mathbf{x}} -
    \hat{L}_{\mathbf{x} - \mathbf{\hat{i}}}\right)^2 + \frac{1}{g^2}
    \sum_{\mathbf{x}} \left(2 -
    \hat{W}_{\mathbf{x}} -\hat{W}^\dagger_{\mathbf{x}} \right) \,.
    \label{eq:u1dual} 
\end{equation}
It can be separated into a free part
\begin{equation}
    \hat{H}_{\textrm{free}}  = 2g^2 \sum_{\mathbf{x}}
    \hat{L}_{\mathbf{x}}^2 + \frac{1}{g^2}
    \sum_{\mathbf{x}} \left(2 -
    \hat{W}_{\mathbf{x}} -\hat{W}^\dagger_{\mathbf{x}} \right)
    \label{eq:free-rotor}
\end{equation}
involving only free or single-site terms,
and a $1$-local or interacting part
\begin{equation}
    \hat{H}_{\textrm{mixing}}  = -g^2 \sum_{\mathbf{x}, i}
    \hat{L}_{\mathbf{x}} \hat{L}_{\mathbf{x} - \mathbf{\hat{i}}}
    \label{eq:mixing-rotor}
\end{equation}
containing only $1$-local terms. The eigenstates of $\hat{H}_{\textrm{free}}$ are well known and prove to be an efficient basis for numerical simulations of the full Hamiltonian
\cite{Jakobs:2025U1,Miranda-Riaza2026}. Crucially, $\hat{H}_{\textrm{free}}$
contains the $\frac{1}{g^2}$ term, that dominates the dynamics at weak couplings.

However, in non-Abelian theories $\hat{L}^c \neq -\hat{R}^c$. Therefore, helper
links remain active, and the parallel transporter $\mathcal{S}_\mathbf{x}$ becomes an up
to $3\ell$-point interaction term for the specific reparametrisation from
Ref.~\cite{PhysRevD.107.074504}. This
poses a significant challenge when it comes to large scale classical and
quantum simulations of the theory and needs to be overcome.

In \cref{sec:static-cosntraints} we will prove that any static constraint
reparametrisation leads to a non-local electric term. In
\cref{sec:dynamic-constraints} we will then explore the possibility of dynamic
constraints. Once again we see limited success, but encounter an
obstruction in non-Abelian theories.
 
\section{Static Constraints}
\label{sec:static-cosntraints}

It is very tempting to keep the number of degrees of freedom to a minimum in any simulation.
This goal naturally favours the use of static constraints (see sec.~\ref{sec:hilbert-space-constraints}) for rotor formulations of the Kogut-Susskind Hamiltonian.
Unfortunately, such a formulation necessarily introduces polynomially non-local many-point terms which typically make simulations infeasible.

We provide a mathematically rigorous statement in \cref{th:non-locality} and the corresponding proof in \cref{sec:formal-proof}.
To put it in a nutshell, the static constraints require the remaining
link variables to form a maximal tree (\cref{th:max-tree}).
Such a tree can always be partitioned into two almost equal parts.
Any interaction between sites corresponding to different parts needs to be mediated along the tree.
Now the number of interacting pairs split into the two parts grows with the system size which means that some of these interactions will be far away from the unique point connecting the parts.
In fact, there will always be neighbouring sites whose distance through the tree grows at least linearly (\cref{th:non-locality}).
Thus, the connection between these sites will be a polynomially non-local many-point term.
For periodic boundary conditions a single link is added to the maximal tree which does not change the polynomial scaling (\cref{th:non-locality-pbc}).

Note that, in principle, this statement is also true for a U$(1)$ gauge theory.
However, in 2D with open boundary conditions all non-local interactions come with exactly zero coupling strength in the Abelian case.
This is a consequence of Gauss' law which reduces to an exactly solvable linear system of equations~\eqref{eq:gauss_law_momentum_u1} in the algebra (momentum basis).
Such a reduction is only possible for Abelian Lie groups and, therefore, the non-local interactions cannot be eliminated in general in the non-Abelian case.

\subsection{Formal statement and proof}\label{sec:formal-proof}
In the following we will use the standard Landau ``big-O-notation''.
In particular, $f\in \Omega(g)$ means that $f$ grows at least as quickly as $g$.

The shortest distance between two points $x,y\in\Lambda$ on a lattice of length $\ell$ differs for open boundary conditions (obc) and periodic boundary conditions (pbc).
To account for this, we define
\begin{align}	
	|x_i-y_i|_\ell &\coloneqq \begin{cases}
		|x_i-y_i| & \text{obc,}\\
		\min\{|x_i-y_i|,\ell -|x_i-y_i|\} & \text{pbc.}
	\end{cases}
\end{align}
We also use the metrics $d_p(x,y) \coloneqq \lVert x-y\rVert_p$ induced by the $p$-norm in the canonical way.
Specifically, we will rely on
\begin{align}
	d_1(x,y) &= \sum_{i=1}^d |x_i-y_i|_\ell\,,\\
	d_\infty(x,y) &= \max_{i=1,\dots, d} |x_i-y_i|_\ell\,,
\end{align}
as well as the tree distance $d_T(x,y)$ defined by the minimal number of links connecting $x$ and $y$ through the tree.
Note the hierarchy
\begin{align}
	d_T(x,y) \ge d_1(x,y) \ge d_\infty(x,y)
\end{align}
for any two points $x,y$.

\begin{lemma}\label{th:max-tree}
  Consider a $d$-dimensional hyper-cubic lattice. 
  \begin{enumerate}
  \item[(a)] For $d=2$ with open boundary conditions, perform a change of variables by introducing rotor degrees of freedom for every plaquette and removing the same number of link variables, that is apply static constraints. Then the remaining helper link variables must form a maximal tree in the original lattice.
  \item[(b)] For $d=2$ with periodic boundary conditions, perform a change of variables by introducing rotor degrees of freedom for every plaquette and removing the same number of link variables, that is apply static constraints. Then the remaining helper link variables must form a maximal tree in the original lattice with exactly one additional link.
  \item[(c)] For $d\ge3$ with open and periodic boundary conditions, there are more plaquettes than link variables that can be removed without disconnecting the lattice.
  Thus, introducing all rotor degrees of freedom cannot be counteracted by removing link degrees of freedom.
  \end{enumerate}
\end{lemma}

\begin{proof}
  Part (a): Let us first count the number of links and plaquettes. Suppose the lattice is of size $m\times n$. Then there are $(m-1)(n-1)$ many lattice sites with links leaving to the right and up. Each of these has a corresponding plaquette. Furthermore, there are $m+n-2$ many boundary vertices on the right and upper end of the lattice. Thus, there are $2(m-1)(n-1)+m+n-2=2mn-m-n$ many links and $(m-1)(n-1)=mn-m-n+1$ many plaquettes. Removing one link for each introduced rotor variable implies that there are $mn-1$ many links remaining.

  Since a tree has exactly one fewer edge than vertices, this directly implies that the remaining helper links either form a maximal tree or disconnect the lattice. However, we cannot disconnect the remaining lattice because that would change the underlying physical system as we could not reconstruct the original configuration which contradicts the (bijective) change of variables.

  Part (b): Following the same approach for periodic boundary conditions, we note that there are $2mn$ many links in the hyper-cubic (rectangular) lattice and $mn$ many plaquettes. Hence, after introducing $mn$ rotor variables and removing as many links, $mn$ many helper links remain. Since we still require a connected helper link graph, $mn-1$ of these links must form a maximal tree with the remaining helper link introducing a single redundant edge. This additional helper link is physically required since the hyper-cubic lattice with periodic boundary conditions contains non-contractible loops that would not be represented by the maximal tree alone. Hence, removing this additional helper link would change the underlying physical system.
  
  Part (c) proceeds similarly by counting the different dofs.
  It is rather technical and conveys little insight.
  We have therefore moved it to \cref{sec:proof_max-tree}.
\end{proof}

However, in this section, we will show that this necessarily leads to non-localities in the electric term by showing that there exist lattice points connected by a single link before the reparametrization that are connected by a non-local path in the tree resulting after reparametrization with static constraints.
Since the connections are mediated by links along the path, it immediately follows that static constraints do not only result in non-locality but also in many-point terms.

More precisely, we will show the following statement.

\begin{theorem}[Non-locality of rotor-Hamiltonians with static constraints]\label{th:non-locality}
	Let $T$ be a spanning tree of a $d$-dimensional hyper-cubic
        lattice $\Lambda=\{1,\ldots,\ell\}^d$ with $d>1$ and open or periodic boundary conditions.
	Then, there exists a pair of neighboring points $x,y\in\Lambda$
        with $d_1(x,y)=1$ which have a tree distance at least linear in
        $\ell$, that is 
	\begin{align}
		\max_{d_1(x,y)=1} d_T(x,y) &\in \Omega\br{\ell}\,.
		\label{eq:lattice_streach}
	\end{align}
\end{theorem}

\begin{proof}
	We first consider periodic and subsequently open boundary conditions.

	\paragraph{Periodic Boundary Conditions:} 
	To show this claim, we first partition the tree into a root $r$ and two sets $A$ and $B$ where $r$ is not a leaf (i.e., it has at least $2$ neighbors in $T$) and every other vertex is either in $A$ or $B$. Furthermore, we enforce that for two vertices $u\in A$ and $v\in B$ the path connecting $u$ and $v$ in $T$ must lead through the root $r$. In other words, if $v$ is a child of $r$, then all descendants of $v$ are in the same set $B$ as $v$. Within the set of all such partitions $(r,A,B)$, there exists a maximal element with respect to the partial order given by the size of the smaller set of $A$ and $B$,i.e.\ $\min(\#A,\#B)$.
	Given the constraint $\#A+\#B=N=\ell^d-1$, this maximal element $(\mathcal{R},\mathcal{A},\mathcal{B})$ is maximally balanced in the sense that the smaller set between $\mathcal{A}$ and $\mathcal{B}$ is maximal.
	
	We can now find bounds on the size of the smaller set of $\mathcal{A}$ and $\mathcal{B}$.
	Without loss of generality, let the smaller set be $\mathcal{B}$.
	
	Case I: Suppose $\mathcal{R}$ has exactly two children $a\in \mathcal{A}$ and $b\in\mathcal{B}$. Since $\mathcal{A}$ is the larger side, $a$ cannot be a leaf, so $(a,\mathcal{A}\setminus\{a\},\mathcal{B}\cup\{\mathcal{R}\})$ is also a valid partition. However, since $\mathcal{B}$ was maximal, $\mathcal{B}\cup\{\mathcal{R}\}$ can no longer be the smaller side. Hence, we conclude $\#\mathcal{B}+1>\#\mathcal{A}-1$, i.e., $\#\mathcal{A}\ge\#\mathcal{B}\ge\#\mathcal{A}-1$ or $\#\mathcal{B}\ge\frac{N-1}{2}$.
	An example of this case is shown in Figure~\ref{fig:Case-I}.
	
	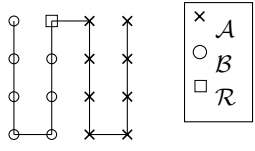
\begin{figure}[t]
		\begin{tikzpicture}[scale = .5,
mycross/.style={
		cross out, 
		draw, 
		minimum size=3pt, 
		line width=0.6pt, 
		inner sep=0pt
	}]
\matrix [draw, fill=white, anchor=north west, xshift=0.5cm] at (3.5,3.5) {
		\node[mycross] {};       & \node[font=\small] {$\mathcal{A}$};  \\
		\node[circle,draw=black, fill=white, inner sep=0pt,minimum size=5pt] {}; & \node[font=\small] {$\mathcal{B}$}; \\
		\node[draw, line width=0.1pt, minimum size=1.5mm, inner sep=0pt] {};      & \node[font=\small] {$\mathcal{R}$}; \\
	};
	\draw (0,3) circle (.13);
	\draw (0,2) circle (.13);
	\draw (0,1) circle (.13);
	\draw (0,0) circle (.13);
	\draw (1,0) circle (.13);
	\draw (1,1) circle (.13);
	\draw (1,2) circle (.13);
	
	\node[draw, line width=0.1pt, minimum size=1.5mm, inner sep=0pt] at (1,3) {};

	\node[mycross] at (2,3) {};
	\node[mycross] at (2,2) {};
	\node[mycross] at (2,1) {};
	\node[mycross] at (2,0) {};
	\node[mycross] at (3,0) {};
	\node[mycross] at (3,1) {};
	\node[mycross] at (3,2) {};
	\node[mycross] at (3,3) {};
	
	\draw (0,0) -- (1,0);
	\draw (1,3) -- (2,3);
	\draw (2,0) -- (3,0);
	
	\draw (0,0) -- (0,1);
	\draw (0,1) -- (0,2);
	\draw (0,2) -- (0,3);
	
	\draw (1,0) -- (1,1);
	\draw (1,1) -- (1,2);
	\draw (1,2) -- (1,3);
	
	\draw (2,0) -- (2,1);
	\draw (2,1) -- (2,2);
	\draw (2,2) -- (2,3);
	
	\draw (3,0) -- (3,1);
	\draw (3,1) -- (3,2);
	\draw (3,2) -- (3,3);
	
\end{tikzpicture} 		\caption{Example tree for Case I on a $4\times 4$ lattice.}
		\label{fig:Case-I}
	\end{figure}
	
	Case II: Suppose $\mathcal{R}$ has exactly $n\ge3$ children $c_1,c_2,\dots,c_n$ and let us denote the descendant set of $c_j$ as $C_j$.
	Each $C_j$ is completely contained in either $\mathcal{A}$ or $\mathcal{B}$.
	Without loss of generality, let $\#C_1\ge\#C_2\ge\cdots \ge\#C_n$.
	If $\#C_1\ge \sum_{j=2}^n\#C_j$, then set $C_1=\mathcal{A}$ and $\mathcal{B}=\cup_{j=2}^n C_j$, and the argument of Case I applies to yield $\#\mathcal{B}\ge\frac{N-1}{2}$.
	An example of this is given in \Cref{fig:case-III-a}.

	Now if $\#C_1 < \sum_{j=2}^n \#C_j$, we define the disjoint sets $D_1\coloneqq C_1$, $D_2\coloneqq \cup_{j\ge1} C_{2j}$, and $D_3\coloneqq \cup_{j\ge1} C_{2j+1}$.
	Their union with the root $\mathcal{R} \cup D_1\cup D_2 \cup D_3$ contains the entire tree.
	Observe that $\#D_3 \le \#D_2$ and $\#D_2 \le \# D_1 + \# D_3$.
	Both inequalities follow from the telescope-like structure: for every set $C_{2j+1}\in D_3$ there is the set $C_{2j}\in D_2$ with $\#C_{2j+1}\le \#C_{2j}$ by construction.
	Similarly, for each $C_{2j}\in D_2$ exists a $C_{2j-1}\in D_1 \cup D_3$ with $\#C_{2j}\le \#C_{2j-1}$.
	Therefore, each of the three sets $D_j$ contains at most half of the $N$ lattice points $\# D_j \le \frac N2$ implying $\# \mathcal{B} \ge \max_j(\#D_j)$.
	Finally, for $N>1$ we can estimate $N=\sum_{j=1}^3 \#D_j \le 3 \max_j (\#D_j) \le 3\# \mathcal{B}$ and, in consequence, $\# \mathcal{B} \ge \frac N3$.

	\begin{figure}[t]
\begin{subfigure}{0.48\textwidth}
			
\begin{tikzpicture}[scale = .5,
mycross/.style={
		cross out, 
		draw, 
		minimum size=3pt, 
		line width=0.6pt, 
		inner sep=0pt
	},
	mystar/.style={
		draw, star, star points=5, star point ratio=2.25, 
		minimum size=4pt, inner sep=0pt
	},
	fullcross/.style={
		path picture={
			\draw[black] (path picture bounding box.south) -- (path picture bounding box.north)
			(path picture bounding box.west) -- (path picture bounding box.east)
			(path picture bounding box.south west) -- (path picture bounding box.north east)
			(path picture bounding box.south east) -- (path picture bounding box.north west);
		},
		inner sep=0pt, minimum size=4.5pt
	},
	mytriangle/.style={
		draw,
		regular polygon, 
		regular polygon sides=3,
		minimum size=4pt, 
		inner sep=0pt
	}
	]
\matrix [draw, fill=white, anchor=north west, xshift=0.5cm] at (3.5,3.5) {
		\node[circle,draw=black, fill=white, inner sep=0pt,minimum size=5pt] {}; & \node[font=\small] {$C_1=\mathcal{A}$}; \\
		\node[mycross] {};       & \node[font=\small] {$C_2$};  \\
		\node[mytriangle] {};       & \node[font=\small] {$C_3$};  \\
		\node[fullcross] {};       & \node[font=\small] {$C_4$};  \\
		\node[draw, line width=0.1pt, minimum size=1.5mm, inner sep=0pt] {};      & \node[font=\small] {$\mathcal{R}$}; \\
	};
	\draw (0,3) circle (.13);
	\draw (0,2) circle (.13);
	\draw (0,1) circle (.13);
\draw (2,3) circle (.13);
	\draw (3,3) circle (.13);
	\draw (3,2) circle (.13);
	\draw (3,1) circle (.13);
	\draw (1,3) circle (.13);
	
	\node[draw, line width=0.1pt, minimum size=1.5mm, inner sep=0pt] at (1,1) {};
	
	\node[fullcross] at (1,2) {};
	
	\node[mycross] at (2,0) {};
	\node[mycross] at (2,1) {};
	\node[mycross] at (2,2) {};
	\node[mycross] at (3,0) {};
	
	\node[mytriangle] at (0,0) {};
	\node[mytriangle] at (1,0) {};
	
	\draw (0,0) -- (1,0);
	\draw (1,1) -- (2,1);
	\draw (2,0) -- (3,0);
	
	\draw (0,1) -- (1,1);
	\draw (0,1) -- (0,2);
	\draw (0,2) -- (0,3);
	
	\draw (1,0) -- (1,1);
	\draw (1,1) -- (1,2);
	\draw (0,3) -- (1,3);
	
	\draw (2,0) -- (2,1);
	\draw (2,1) -- (2,2);
	\draw (1,3) -- (2,3);
	
	\draw (3,1) -- (3,2);
	\draw (2,3) -- (3,3);
	\draw (3,3) -- (3,2);
	
\end{tikzpicture} 			\caption{$\#C_1\geq \#C_2+\#C_3+\#C_4$ and $C_2\cup C_3\cup C_4=\mathcal{B}$.}
			\label{fig:case-III-a}
		\end{subfigure}
\begin{subfigure}{0.48\textwidth}
			
\begin{tikzpicture}[scale = .5,
mycross/.style={
		cross out, 
		draw, 
		minimum size=3pt, 
		line width=0.6pt, 
		inner sep=0pt
	},
	mystar/.style={
		draw, star, star points=5, star point ratio=2.25, 
		minimum size=4pt, inner sep=0pt
	},
	fullcross/.style={
		path picture={
			\draw[black] (path picture bounding box.south) -- (path picture bounding box.north)
			(path picture bounding box.west) -- (path picture bounding box.east)
			(path picture bounding box.south west) -- (path picture bounding box.north east)
			(path picture bounding box.south east) -- (path picture bounding box.north west);
		},
		inner sep=0pt, minimum size=4.5pt
	},
	mytriangle/.style={
		draw,
		regular polygon, 
		regular polygon sides=3,
		minimum size=4pt, 
		inner sep=0pt
	}
	]
\matrix [draw, fill=white, anchor=north west, xshift=0.5cm] at (3.5,3.5) {
		\node[circle,draw=black, fill=white, inner sep=0pt,minimum size=5pt] {}; & \node[font=\small] {$C_1=\mathcal{B}$}; \\
		\node[mycross] {};       & \node[font=\small] {$C_2$};  \\
		\node[mytriangle] {};       & \node[font=\small] {$C_3$};  \\
\node[draw, line width=0.1pt, minimum size=1.5mm, inner sep=0pt] {};      & \node[font=\small] {$\mathcal{R}$}; \\
	};
	\draw (0,3) circle (.13);
	\draw (0,2) circle (.13);
	\draw (2,3) circle (.13);
	\draw (3,3) circle (.13);
	\draw (1,3) circle (.13);
	
	\node[draw, line width=0.1pt, minimum size=1.5mm, inner sep=0pt] at (1,2) {};
	
	\node[mycross] at (3,2) {};
	\node[mycross] at (2,1) {};
	\node[mycross] at (2,2) {};
	\node[mycross] at (3,0) {};
	\node[mycross] at (3,1) {};
	
	\node[mytriangle] at (0,0) {};
	\node[mytriangle] at (1,0) {};
	\node[mytriangle] at (2,0) {};
	\node[mytriangle] at (0,1) {};
	\node[mytriangle] at (1,1) {};
	
	\draw (0,0) -- (1,0);
	\draw (1,0) -- (2,0);
	\draw (2,2) -- (3,2);
	
	\draw (0,1) -- (1,1);
	\draw (1,2) -- (0,2);
	\draw (0,2) -- (0,3);
	
	\draw (1,0) -- (1,1);
	\draw (1,1) -- (1,2);
	\draw (0,3) -- (1,3);
	
	\draw (3,0) -- (3,1);
	\draw (2,1) -- (2,2);
	\draw (1,3) -- (2,3);
	
	\draw (3,1) -- (2,1);
	\draw (2,3) -- (3,3);
	\draw (1,2) -- (2,2);
	
\end{tikzpicture} 			\caption{$\#C_1< \#C_2+\#C_3$, $C_2\cup C_3=\mathcal{A}$ and $\#\mathcal{B}=5=\frac N3$.}
			\label{fig:case-III-b}
		\end{subfigure}
\begin{subfigure}{0.48\textwidth}
			
\begin{tikzpicture}[scale = .5,
mycross/.style={
		cross out, 
		draw, 
		minimum size=3pt, 
		line width=0.6pt, 
		inner sep=0pt
	},
	mystar/.style={
		draw, star, star points=5, star point ratio=2.25, 
		minimum size=4pt, inner sep=0pt
	},
	fullcross/.style={
		path picture={
			\draw[black] (path picture bounding box.south) -- (path picture bounding box.north)
			(path picture bounding box.west) -- (path picture bounding box.east)
			(path picture bounding box.south west) -- (path picture bounding box.north east)
			(path picture bounding box.south east) -- (path picture bounding box.north west);
		},
		inner sep=0pt, minimum size=4.5pt
	},
	mytriangle/.style={
		draw,
		regular polygon, 
		regular polygon sides=3,
		minimum size=4pt, 
		inner sep=0pt
	}
	]
\matrix [draw, fill=white, anchor=north west, xshift=0.5cm] at (3.5,3.5) {
		\node[circle,draw=black, fill=white, inner sep=0pt,minimum size=5pt] {}; & \node[font=\small] {$C_1$}; \\
		\node[mycross] {};       & \node[font=\small] {$C_2$};  \\
		\node[mytriangle] {};       & \node[font=\small] {$C_3$};  \\
		\node[fullcross] {};       & \node[font=\small] {$C_4$};  \\
		\node[draw, line width=0.1pt, minimum size=1.5mm, inner sep=0pt] {};      & \node[font=\small] {$\mathcal{R}$}; \\
	};
	\draw (0,3) circle (.13);
	\draw (0,2) circle (.13);
	\draw (0,1) circle (.13);
	\draw (2,3) circle (.13);
	\draw (3,3) circle (.13);
\draw (1,3) circle (.13);
	
	\node[draw, line width=0.1pt, minimum size=1.5mm, inner sep=0pt] at (1,1) {};
	
	\node[fullcross] at (1,2) {};
	
	\node[mytriangle] at (2,0) {};
	\node[mycross] at (3,2) {};
	\node[mycross] at (2,1) {};
	\node[mycross] at (2,2) {};
	\node[mycross] at (3,0) {};
	\node[mycross] at (3,1) {};
	
	\node[mytriangle] at (0,0) {};
	\node[mytriangle] at (1,0) {};
	
	\draw (0,0) -- (1,0);
	\draw (1,1) -- (2,1);
	\draw (2,1) -- (3,1);
	
	\draw (0,1) -- (1,1);
	\draw (0,1) -- (0,2);
	\draw (0,2) -- (0,3);
	
	\draw (1,0) -- (1,1);
	\draw (1,1) -- (1,2);
	\draw (0,3) -- (1,3);
	
	\draw (1,0) -- (2,0);
	\draw (2,1) -- (2,2);
	\draw (1,3) -- (2,3);
	
	\draw (3,1) -- (3,0);
	\draw (2,3) -- (3,3);
	\draw (3,2) -- (3,1);
	
\end{tikzpicture} 			\caption{$\#C_1 < \#C_2+\#C_3+\#C_4$ and $\mathcal{B}=C_1\cup C_4$. }
			\label{fig:case-III-c}
		\end{subfigure}
		\caption{Example trees for Case II on a $4\times 4$ lattice.}
		\label{fig:Case-II}
	\end{figure}
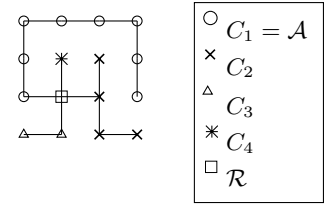
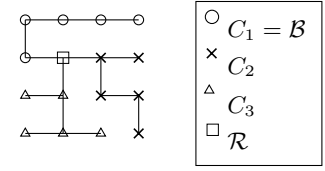
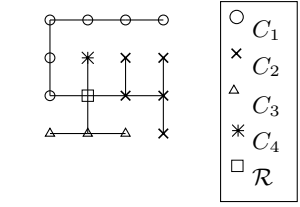
	
	While not relevant for the proof, we note that this bound is tight since e.g.\ for the tree in \Cref{fig:case-III-b} the bound $\#\mathcal{B}=5=\frac N3$ is saturated.
	The construction itself does not always lead to the best choice of $\mathcal{B}$ as can be seen in \Cref{fig:case-III-c} where $\#D_1=\#D_2=6$ but $\#\mathcal{B}=\#C_1+\#C_4=7$.
	
	With periodic boundary conditions (pbc) in $d$ dimensions every point in $\mathcal{B}$ has either all $2d$ neighbours from $\mathcal{B}$ (interior) or at least one neighbour in $\mathcal{A} \cup \mathcal{R}$ (boundary).
	There are at most $2d-1$ points in $\mathcal{B}$ neighbouring $\mathcal{R}$ for any $\ell$, so the entire remaining boundary $\partial \mathcal{B}$ must be adjacent to $\mathcal{A}$.
	
	If there is a point $x\in\partial \mathcal{B}$ with tree distance from the root $d_T(x,\mathcal{R})\ge \lfloor \frac\ell 2\rfloor$, then the claim of \cref{th:non-locality} is already fulfilled.
	Thus, we focus on the case when for all $x\in\partial \mathcal{B}$ we have $\lfloor \frac\ell 2\rfloor > d_T(x,\mathcal{R}) \ge d_\infty(x,\mathcal{R})$ and therefore the path from the root to the boundary is always shorter ``through the tree'' rather than wrapping around via the pbc.
	This implies that there is no entire $(d-1)$-dimensional hyperplane of points $z\in\mathcal{B} \setminus \partial \mathcal{B}$ with $d_\infty(z,\mathcal{R})=\lfloor \frac\ell 2\rfloor$ since otherwise the tree distance to the boundary on the other side of this hyperplane would be larger than $\lfloor \frac\ell 2\rfloor$.
	Thus, for every $x$ in the interior of $\mathcal{B}$, there exists a point $y\in \mathcal{B}$ with distance $d_\infty(y,R)=d_\infty(x,R)+1$.
	Hence, the maximal $d_\infty$-distance points in $\mathcal{B}$ must lie on the boundary $\partial \mathcal{B}$.

	There are $N_\delta=(2\delta+1)^d-1$ vertices with $\infty$-distance at most $\delta$ from the root in the hyper-cubic lattice.
If $\# \mathcal B>N_{d}$, then there is a point in $\mathcal{B}$ with distance from the root larger than $\delta$ and therefore a point adjacent to $\mathcal{A}$ with distance at least $\delta$.
	Finally, $\#\mathcal{B} \ge \frac N3 > N_{\delta}$ unless $\delta\ge\frac12 \br{\br{\frac N3+1}^{1/d}-1}\in\Theta(\ell)$.

	Since the tree distance $d_T$ is at least as large as the distance $d_\infty$, this means that there exists a vertex $b\in\mathcal{B}$ that neighbors a vertex $a\in\mathcal{A}$ prior to reparametrization whose distance to the root is $\Omega({\ell})$.
	Since the path from $b$ to $a$ in $T$ must go through the root, we have found a pair of vertices that were connected by a single link prior to reparametrization, which are now connected by a path of length $\Omega(\ell)$.
	
	This completes the proof that static constraints lead to non-local interactions in the electric term for periodic boundary conditions.
	\paragraph{Open Boundary Condition:}
	The proof of this case can be found in ~\cite{lin2018minimumtreestretchhamminggraphs}, where it was shown that 
	\begin{equation}
		\max_{d_1(x,y)=1} d_T(x,y)\ge 2(d-1)\left\lfloor\frac{\ell}{2}\right\rfloor+1 \,,
	\end{equation}
	which implies \cref{eq:lattice_streach}.
\end{proof}

\Cref{th:non-locality} is not entirely sufficient to show non-locality for pbc because in that case there is one additional link introducing a loop on top of the maximal tree.
Below we show that the minimally linear non-locality generalises for any constant number $K$ of additional links and thus in particular for the case $K=1$ of interest.

\begin{corollary}\label{th:non-locality-pbc}
	Let $T$ be the union of a spanning tree and $K$ additional links of a $d$-dimensional hyper-cubic
	lattice $\Lambda=\{1,\ldots,\ell\}^d$ with $d>1$ and periodic boundary conditions.
	Then, there exists a pair of neighboring points $x,y\in\Lambda$
	with $d_1(x,y)=1$ which have a tree distance at least linear in
	$\ell$, more specifically
	\begin{align}
		\max_{d_1(x,y)=1} d_T(x,y) &\in \Omega\br{\frac{\ell}{(K+1)^{1/d}}}\,.
	\end{align}
\end{corollary}

\begin{proof}
  Let $T_0$ be a maximal tree in $T$, i.e., $T_0$ and $T$ coincide with the exception of the $K$ additional links in $T$. As in the proof of \cref{th:non-locality}, we can find a maximal partition $(\mathcal{R},\mathcal{A},\mathcal{B})$ of $T_0$ and assume $\mathcal{B}$ to be the smaller set. This partition also applies to $T$ except that the additional links may introduce ``shortcuts'' between $\mathcal{A}$ and $\mathcal{B}$. First however, we note that the distances $d_T$ and $d_{T_0}$, in $T$ and $T_0$ respectively, satisfy $d_{T_0}\ge d_T\ge d_\infty$. Thus, for all additional links connecting vertices that are both in $\mathcal{A}$ or both in $\mathcal{B}$ the proof of \cref{th:non-locality} holds verbatim and these do not contribute shortcuts.

  Let us therefore now assume the ``worst case'', namely that all the additional links in $T$ have one end $u_i\in\mathcal{A}$ and the other end $v_i\in\mathcal{B}$, $i\in\brc{1,\dots,K}$. Since $T_0$ is a tree, this implies that the $i$-th additional link closes a cycle if added to $T_0$. The only other paths connecting $u_i$ and $v_i$ must contain more shortcuts through some other additional links or pass through the root $\mathcal{R}$ because $u_i$ is a descendant of a root neighbor $u_i'\in\mathcal{A}$ and $v_i$ is a descendant of a root neighbor $v_i'\in\mathcal{B}$. Hence, removing the additional links as well as $\mathcal{R}$ from any path in $T$ leaves at most $K+1$ many paths, each of which are fully contained in $\mathcal{A}$ or $\mathcal{B}$ because each link in the remaining sub-paths connects a parent and a child in $T_0$ which both must lie in the same set $\mathcal{A}$ or $\mathcal{B}$. Any point $x\in\partial\mathcal{B}$ therefore connects to its neighbor in $\mathcal{A}$ either via $\mathcal{R}$ or one of the $u_i$.

  We can now return to the argument in the proof of \cref{th:non-locality} noting that the number of points with $\infty$-distance at most $\delta$ from either $\mathcal{R}$ or any $u_i$ is $(K+1)N_\delta$ and thus $\#\mathcal{B} \ge \frac N3 > (K+1)N_{\delta}$ unless $\delta\ge\frac12 \br{\br{\frac N{3(K+1)}+1}^{1/d}-1}\in\Theta\br{\frac{\ell}{(K+1)^{1/d}}}$.
\end{proof}

\subsection{Rotor vs.\ locality trade-off}
While it is not possible to introduce a rotor variable for every plaquette and maintain locality with static constraints, we can introduce a trade-off where we introduce fewer rotor variables than plaquettes to ensure predictable locality scaling.

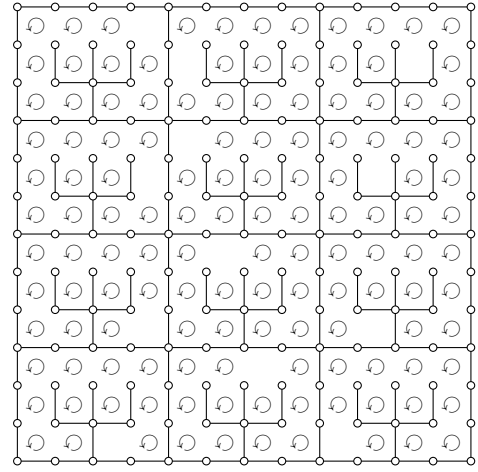
\begin{figure}
	
\begin{tikzpicture}[scale = .5]
	\foreach \j in {0,...,12}{
		\foreach \k in {0,...,12}{
			\draw (\j,\k) circle (.1);
		}
	}
	\foreach \j in {0,...,3}{
		\foreach \k in {0,...,11}{
			\draw (4*\j,\k+.1) -- (4*\j,\k+.9);
		}
	}
	\foreach \j in {0,...,4}{
		\foreach \k in {0,...,11}{
			\draw (\k+.1,3*\j) -- (\k+.9,3*\j);
		}
	}
	\foreach \j in {0,...,2}{
		\foreach \k in {0,...,3}{
			\draw (4*\j+2,3*\k+.1)--(4*\j+2,3*\k+.9);
			\draw (4*\j+1,3*\k+1.1)--(4*\j+1,3*\k+1.9);
			\draw (4*\j+2,3*\k+1.1)--(4*\j+2,3*\k+1.9);
			\draw (4*\j+3,3*\k+1.1)--(4*\j+3,3*\k+1.9);
			\draw (4*\j+1.1,3*\k+1)--(4*\j+1.9,3*\k+1);
			\draw (4*\j+2.1,3*\k+1)--(4*\j+2.9,3*\k+1);
		}
	}
\node[rotate=135] (note) at (1.5,.5) {$\circlearrowleft$};
	\node[rotate=135] (note) at (.5,.5) {$\circlearrowleft$};
	\node[rotate=135] (note) at (.5,1.5) {$\circlearrowleft$};
	\node[rotate=135] (note) at (.5,2.5) {$\circlearrowleft$};
	\node[rotate=135] (note) at (1.5,2.5) {$\circlearrowleft$};
	\node[rotate=135] (note) at (2.5,2.5) {$\circlearrowleft$};
	\node[rotate=135] (note) at (3.5,2.5) {$\circlearrowleft$};
	\node[rotate=135] (note) at (3.5,1.5) {$\circlearrowleft$};
	\node[rotate=135] (note) at (3.5,.5) {$\circlearrowleft$};
	\node[rotate=135] (note) at (1.5,1.5) {$\circlearrowleft$};
	\node[rotate=135] (note) at (2.5,1.5) {$\circlearrowleft$};
\node[rotate=135] (note) at (1.5,3.5) {$\circlearrowleft$};
	\node[rotate=135] (note) at (.5,3.5) {$\circlearrowleft$};
	\node[rotate=135] (note) at (.5,4.5) {$\circlearrowleft$};
	\node[rotate=135] (note) at (.5,5.5) {$\circlearrowleft$};
	\node[rotate=135] (note) at (1.5,5.5) {$\circlearrowleft$};
	\node[rotate=135] (note) at (2.5,5.5) {$\circlearrowleft$};
	\node[rotate=135] (note) at (3.5,5.5) {$\circlearrowleft$};
	\node[rotate=135] (note) at (3.5,4.5) {$\circlearrowleft$};
	\node[rotate=135] (note) at (2.5,3.5) {$\circlearrowleft$};
	\node[rotate=135] (note) at (1.5,4.5) {$\circlearrowleft$};
	\node[rotate=135] (note) at (2.5,4.5) {$\circlearrowleft$};
\node[rotate=135] (note) at (1.5,6.5) {$\circlearrowleft$};
	\node[rotate=135] (note) at (.5,6.5) {$\circlearrowleft$};
	\node[rotate=135] (note) at (.5,7.5) {$\circlearrowleft$};
	\node[rotate=135] (note) at (.5,8.5) {$\circlearrowleft$};
	\node[rotate=135] (note) at (1.5,8.5) {$\circlearrowleft$};
	\node[rotate=135] (note) at (2.5,8.5) {$\circlearrowleft$};
	\node[rotate=135] (note) at (3.5,8.5) {$\circlearrowleft$};
	\node[rotate=135] (note) at (3.5,6.5) {$\circlearrowleft$};
	\node[rotate=135] (note) at (2.5,6.5) {$\circlearrowleft$};
	\node[rotate=135] (note) at (1.5,7.5) {$\circlearrowleft$};
	\node[rotate=135] (note) at (2.5,7.5) {$\circlearrowleft$};
\node[rotate=135] (note) at (1.5,9.5) {$\circlearrowleft$};
	\node[rotate=135] (note) at (.5,9.5) {$\circlearrowleft$};
	\node[rotate=135] (note) at (.5,10.5) {$\circlearrowleft$};
	\node[rotate=135] (note) at (.5,11.5) {$\circlearrowleft$};
	\node[rotate=135] (note) at (1.5,11.5) {$\circlearrowleft$};
	\node[rotate=135] (note) at (2.5,11.5) {$\circlearrowleft$};
	\node[rotate=135] (note) at (3.5,10.5) {$\circlearrowleft$};
	\node[rotate=135] (note) at (3.5,9.5) {$\circlearrowleft$};
	\node[rotate=135] (note) at (2.5,9.5) {$\circlearrowleft$};
	\node[rotate=135] (note) at (1.5,10.5) {$\circlearrowleft$};
	\node[rotate=135] (note) at (2.5,10.5) {$\circlearrowleft$};
\node[rotate=135] (note) at (5.5,.5) {$\circlearrowleft$};
	\node[rotate=135] (note) at (4.5,.5) {$\circlearrowleft$};
	\node[rotate=135] (note) at (4.5,1.5) {$\circlearrowleft$};
	\node[rotate=135] (note) at (4.5,2.5) {$\circlearrowleft$};
	\node[rotate=135] (note) at (5.5,2.5) {$\circlearrowleft$};
	\node[rotate=135] (note) at (6.5,.5) {$\circlearrowleft$};
	\node[rotate=135] (note) at (7.5,2.5) {$\circlearrowleft$};
	\node[rotate=135] (note) at (7.5,1.5) {$\circlearrowleft$};
	\node[rotate=135] (note) at (7.5,.5) {$\circlearrowleft$};
	\node[rotate=135] (note) at (5.5,1.5) {$\circlearrowleft$};
	\node[rotate=135] (note) at (6.5,1.5) {$\circlearrowleft$};
\node[rotate=135] (note) at (5.5,3.5) {$\circlearrowleft$};
	\node[rotate=135] (note) at (4.5,3.5) {$\circlearrowleft$};
	\node[rotate=135] (note) at (4.5,4.5) {$\circlearrowleft$};
	\node[rotate=135] (note) at (4.5,5.5) {$\circlearrowleft$};
	\node[rotate=135] (note) at (6.5,5.5) {$\circlearrowleft$};
	\node[rotate=135] (note) at (6.5,3.5) {$\circlearrowleft$};
	\node[rotate=135] (note) at (7.5,5.5) {$\circlearrowleft$};
	\node[rotate=135] (note) at (7.5,4.5) {$\circlearrowleft$};
	\node[rotate=135] (note) at (7.5,3.5) {$\circlearrowleft$};
	\node[rotate=135] (note) at (5.5,4.5) {$\circlearrowleft$};
	\node[rotate=135] (note) at (6.5,4.5) {$\circlearrowleft$};
\node[rotate=135] (note) at (5.5,6.5) {$\circlearrowleft$};
	\node[rotate=135] (note) at (4.5,6.5) {$\circlearrowleft$};
	\node[rotate=135] (note) at (4.5,7.5) {$\circlearrowleft$};
	\node[rotate=135] (note) at (5.5,8.5) {$\circlearrowleft$};
	\node[rotate=135] (note) at (6.5,8.5) {$\circlearrowleft$};
	\node[rotate=135] (note) at (6.5,6.5) {$\circlearrowleft$};
	\node[rotate=135] (note) at (7.5,8.5) {$\circlearrowleft$};
	\node[rotate=135] (note) at (7.5,7.5) {$\circlearrowleft$};
	\node[rotate=135] (note) at (7.5,6.5) {$\circlearrowleft$};
	\node[rotate=135] (note) at (5.5,7.5) {$\circlearrowleft$};
	\node[rotate=135] (note) at (6.5,7.5) {$\circlearrowleft$};
\node[rotate=135] (note) at (5.5,9.5) {$\circlearrowleft$};
	\node[rotate=135] (note) at (4.5,9.5) {$\circlearrowleft$};
	\node[rotate=135] (note) at (4.5,11.5) {$\circlearrowleft$};
	\node[rotate=135] (note) at (5.5,11.5) {$\circlearrowleft$};
	\node[rotate=135] (note) at (6.5,11.5) {$\circlearrowleft$};
	\node[rotate=135] (note) at (6.5,9.5) {$\circlearrowleft$};
	\node[rotate=135] (note) at (7.5,11.5) {$\circlearrowleft$};
	\node[rotate=135] (note) at (7.5,10.5) {$\circlearrowleft$};
	\node[rotate=135] (note) at (7.5,9.5) {$\circlearrowleft$};
	\node[rotate=135] (note) at (5.5,10.5) {$\circlearrowleft$};
	\node[rotate=135] (note) at (6.5,10.5) {$\circlearrowleft$};
\node[rotate=135] (note) at (9.5,.5) {$\circlearrowleft$};
	\node[rotate=135] (note) at (10.5,2.5) {$\circlearrowleft$};
	\node[rotate=135] (note) at (8.5,1.5) {$\circlearrowleft$};
	\node[rotate=135] (note) at (8.5,2.5) {$\circlearrowleft$};
	\node[rotate=135] (note) at (9.5,2.5) {$\circlearrowleft$};
	\node[rotate=135] (note) at (10.5,.5) {$\circlearrowleft$};
	\node[rotate=135] (note) at (11.5,2.5) {$\circlearrowleft$};
	\node[rotate=135] (note) at (11.5,1.5) {$\circlearrowleft$};
	\node[rotate=135] (note) at (11.5,.5) {$\circlearrowleft$};
	\node[rotate=135] (note) at (9.5,1.5) {$\circlearrowleft$};
	\node[rotate=135] (note) at (10.5,1.5) {$\circlearrowleft$};
\node[rotate=135] (note) at (8.5,3.5) {$\circlearrowleft$};
	\node[rotate=135] (note) at (10.5,5.5) {$\circlearrowleft$};
	\node[rotate=135] (note) at (8.5,4.5) {$\circlearrowleft$};
	\node[rotate=135] (note) at (8.5,5.5) {$\circlearrowleft$};
	\node[rotate=135] (note) at (9.5,5.5) {$\circlearrowleft$};
	\node[rotate=135] (note) at (10.5,3.5) {$\circlearrowleft$};
	\node[rotate=135] (note) at (11.5,5.5) {$\circlearrowleft$};
	\node[rotate=135] (note) at (11.5,4.5) {$\circlearrowleft$};
	\node[rotate=135] (note) at (11.5,3.5) {$\circlearrowleft$};
	\node[rotate=135] (note) at (9.5,4.5) {$\circlearrowleft$};
	\node[rotate=135] (note) at (10.5,4.5) {$\circlearrowleft$};
\node[rotate=135] (note) at (8.5,6.5) {$\circlearrowleft$};
	\node[rotate=135] (note) at (10.5,8.5) {$\circlearrowleft$};
	\node[rotate=135] (note) at (8.5,7.5) {$\circlearrowleft$};
	\node[rotate=135] (note) at (8.5,8.5) {$\circlearrowleft$};
	\node[rotate=135] (note) at (9.5,8.5) {$\circlearrowleft$};
	\node[rotate=135] (note) at (10.5,6.5) {$\circlearrowleft$};
	\node[rotate=135] (note) at (11.5,8.5) {$\circlearrowleft$};
	\node[rotate=135] (note) at (11.5,7.5) {$\circlearrowleft$};
	\node[rotate=135] (note) at (11.5,6.5) {$\circlearrowleft$};
	\node[rotate=135] (note) at (9.5,6.5) {$\circlearrowleft$};
	\node[rotate=135] (note) at (10.5,7.5) {$\circlearrowleft$};
\node[rotate=135] (note) at (8.5,9.5) {$\circlearrowleft$};
	\node[rotate=135] (note) at (10.5,11.5) {$\circlearrowleft$};
	\node[rotate=135] (note) at (8.5,10.5) {$\circlearrowleft$};
	\node[rotate=135] (note) at (8.5,11.5) {$\circlearrowleft$};
	\node[rotate=135] (note) at (9.5,11.5) {$\circlearrowleft$};
	\node[rotate=135] (note) at (10.5,9.5) {$\circlearrowleft$};
	\node[rotate=135] (note) at (11.5,11.5) {$\circlearrowleft$};
	\node[rotate=135] (note) at (11.5,10.5) {$\circlearrowleft$};
	\node[rotate=135] (note) at (11.5,9.5) {$\circlearrowleft$};
	\node[rotate=135] (note) at (9.5,9.5) {$\circlearrowleft$};
	\node[rotate=135] (note) at (9.5,10.5) {$\circlearrowleft$};
\end{tikzpicture}   \caption{Example of a $12\times12$-plaquette system decomposed into $3\times 4$-plaquette cells with remaining edges. The cell boundary contains all original links. The interior is a maximal tree connected by one additional link to the boundary. Each cell has $11$ broken links, i.e., we can introduce $11$ of the maximally $12$ rotor variables. Each $3\times 4$-plaquette cell shows one of the $12$ choices.}\label{fig:rotor-locality-example}
\end{figure}

As illustrated in \Cref{fig:rotor-locality-example}, we can sub-divide a $M\times N$-plaquette system into $m\times n$-plaquette cells that are fully connected on the boundary and reduced to a maximal tree in the interior. In doing so, any pair of neighboring vertices prior to reparametrization is contained in one such $m\times n$-plaquette cell. Furthermore, any two vertices in a cell have a distance of no more than $m+n+\mathrm{diam}(\text{internal tree})+1$. Since trees have exactly one fewer edge than vertices, and an $m\times n$-plaquette cell has exactly $(m-1)(n-1)$ many internal vertices, we obtain an upper bound on the non-locality given by $mn+1$.

To obtain the ratio of rotor variables that can be introduced, we note that an $m\times n$-plaquette system has a total of $2mn+m+n$ many links. Since the internal vertices form an $(m-2)\times(n-2)$-plaquette system, there are $2mn-3m-3n+4$ links not touching a vertex on the boundary. Furthermore, there are $2(m-1)+2(n-1)$ many links connecting a boundary vertex to an internal vertex. Thus, there are a total of $2mn-m-n$ many non-boundary links. Of these, the internal tree plus one many remain, i.e., $(m-1)(n-1)$ many links are remaining. In other words, we can break up to $mn-1$ non-boundary links. Since each broken link allows for exactly one rotor variable to be introduced, the fraction of rotor variables that can be introduced with this tiling is $\frac{mn-1}{mn}$ while maintaining a locality bound of $mn+1$.

However, this trade-off is likely not useful for numerical
simulations, as the remaining plaquette will be expressed as a many-point term consisting of links and rotors found within the cell. Therefore, they will exhibit similar numerical problems as encountered when simulating the original Kogut-Susskind Hamiltonian.

\section{Dynamic Constraints}
\label{sec:dynamic-constraints}

Another route to a local system might be to introduce more gauge degrees of
freedom alongside additional constraints. This idea was first proposed in
\cite{PhysRevD.107.074504}. However, it is not clear how the constraint
proposed there can be implemented for numerical simulations.

Nevertheless, the overall approach seems to show some promise. In the following
we will present a plaquette separation ansatz and derive the correct dual
Hamiltonian for U(1). Then we attempt a similar formulation in SU(2) and show
that it breaks.

\subsection{Dynamic Constraints in U(1)}
\label{sec:dyn-constraints-U1}

To begin with we again look at a pure U(1) lattice gauge theory in two spatial
dimensions with open boundary conditions. The gauge links will be parametrised
as
\begin{equation}
	U_{\mathbf{x},i} (\varphi) = \mathrm{e}^{\im \varphi_{\mathbf{x},i}} \,
	.
\end{equation}
Then the Kogut-Susskind Hamiltonian reads
\begin{equation}
  \begin{split}
    \hat{H} = - \frac{g^2}{2} & \sum_{\mathbf{x},i} 
    \frac{\md^2}{\md
      \varphi_{\mathbf{x},i}^2} +
    \frac{2}{g^2} \sum_{\mathbf{x}} \left( 1 \right. \\
    & - \cos\left( \varphi_{\mathbf{x},1} +
    \varphi_{\mathbf{x}+\hat{\mathbf{1}},2}
    \left. - \varphi_{\mathbf{x}+\hat{\mathbf{2}},1} -
    \varphi_{\mathbf{x},2}	\right)  \right) \, .
  \end{split}
\end{equation}
To construct a dual Hamiltonian we begin by introducing additional independent
plaquette degrees of freedom
\begin{equation}
	W_{\mathbf{x}} = \mathrm{e}^{\im \omega_{\mathbf{x}}}\,.
\end{equation}
The goal is for these dof to capture all excitations of the plaquette
found at position $\mathbf{x}$. I.e.\ the magnetic Hamiltonian simplifies to
\begin{equation}
	\hat{H}_{\textrm{magnetic}} = \frac{2}{g^2} \sum_{\mathbf{x}} (1 - \cos
	\omega_{\mathbf{x}}) \, .
\end{equation}
As discussed earlier, such a locally acting magnetic term is essential for
efficient numerical simulations at weak couplings.

Of course just introducing new degrees of freedom will change the physics of
the system, if no counter measures are taken. In particular, our wave function
is now	not just a function of the links, but also the new plaquette dof:
\begin{equation}
	\psi(\dots, U_{\mathbf{x},i},\dots) \rightarrow
	\widetilde{\psi}(\dots, \widetilde{U}_{\mathbf{x},i},\dots,
	\widetilde{W}_{\mathbf{x}}, \dots) \, .
\end{equation}
Operators and variables in this enlarged Hilbert space will be denoted with a
$\widetilde{\,\,}$ in the following. The gauge links
$\widetilde{U}_{\mathbf{x},i}$ will be referred to as helper links as before.

In order to obtain the same physics in this enlarged Hilbert space
additional constraints are needed. In particular, we want to prohibit any
plaquette excitations in the helper links
$\widetilde{U}_{\mathbf{x},i}$. Or put differently we want the wave function to
be independent of the value of the (helper link) plaquette
\begin{align}
	\widetilde{P}_{\mathbf{x}} & = \widetilde{U}_{\mathbf{x},i}\,
	\widetilde{U}_{\mathbf{x}+\hat{\mathbf{i}}, j}\,
	\widetilde{U}^\dagger_{\mathbf{x}+\hat{\mathbf{j}}, i}\,
	\widetilde{U}^\dagger_{\mathbf{x},j}                          \\
	                           & = \mathrm{e}^{i \left(
	\widetilde{\varphi}_{\mathbf{x},i} +
	\widetilde{\varphi}_{\mathbf{x}+\hat{\mathbf{i}}, j}
	- \widetilde{\varphi}_{\mathbf{x}+\hat{\mathbf{j}}, i}
	- \widetilde{\varphi}_{\mathbf{x},j}  \right)} = \mathrm{e}^{i
	\widetilde{\varphi}^P_{\mathbf{x}}} \, .
\end{align}
Thus we demand
\begin{align}
	\frac{\partial}{\partial \widetilde{\varphi}^P_{\mathbf{x}}} \,
	\widetilde{\psi} & = \sum_{\mathbf{x}, i} \frac{\partial
	\widetilde{\varphi}_{\mathbf{x},i}}{\partial
	\widetilde{\varphi}^P_{\mathbf{x}}}
	\frac{\partial \widetilde{\psi}}{\partial
	\widetilde{\varphi}_{\mathbf{x},i}}                      \\
	\begin{split}
		& = \left(
		\frac{\partial}{\partial
			\widetilde{\varphi}_{\mathbf{x},i}} +
		\frac{\partial}{\partial
			\widetilde{\varphi}_{\mathbf{x}+\hat{\mathbf{i}}, j}}
		\right.
		\\
		& \left. \qquad \qquad - \frac{\partial}{\partial
			\widetilde{\varphi}_{\mathbf{x}+\hat{\mathbf{j}}, i}}
		- \frac{\partial}{\partial
			\widetilde{\varphi}_{\mathbf{x},j}} \right)
		\widetilde{\psi}
		= 0 \, .
	\end{split}
	\label{eq:u1-plaq-constraint}
\end{align}
To get an intuition of what this constraint means in practice we can look at
its action on the electric basis states. The electric basis is spanned by the
basis functions:
\begin{equation}
	\ket{\dots, m_{\mathbf{x}, i}, \dots} = \prod_{\mathbf{x}, i}
	\frac{\mathrm{e}^{\im m_{\mathbf{x}, i} \,
	\widetilde{\varphi}_{\mathbf{x},
	i}}}{\sqrt{2 \pi}} \quad , \, m_{\mathbf{x}, i} \in \mathbb{Z}
\end{equation}
States fulfilling the plaquette constraint thus must obey
\begin{equation}
	m_{\mathbf{x},i} +
	m_{\mathbf{x}+\hat{\mathbf{i}}, j}
	- m_{\mathbf{x}+\hat{\mathbf{j}}, i}
	- m_{\mathbf{x},j} = 0 \, .
	\label{eq:plaquette_constraints}
\end{equation}
This offers a neat physical interpretation: the plaquette constraint simply
demands that the oriented sum of charges of each plaquette vanishes.

The next step is to find appropriate electric operators acting on
$\widetilde{\psi}$. The electric charge on the original link $U_{\mathbf{x},i}$
is split up between the helper link $\widetilde{U}_{\mathbf{x},i}$ and the
rotors $\widetilde{W}_{\mathbf{x}}$ and $\widetilde{W}_{\mathbf{x} -
	\hat{\mathbf{i}}}$. To calculate its exact shape we can consider the
map from
the dual wave function $\widetilde{\psi}$ back to a link wave function $\psi$.
It is given by
\begin{equation}
	\begin{split}
		\psi \left( \dots, U_{\mathbf{x},i}, \dots\right) &=
		\widetilde{\psi} \left(\dots,\widetilde{U}_{\mathbf{x},i} =
		U_{\mathbf{x},i},
		\dots, \right. \\
		& \left. \widetilde{W}_{\mathbf{x}} = U_{\mathbf{x},i}\,
		U_{\mathbf{x}+\hat{\mathbf{i}}, j}\,
		U^\dagger_{\mathbf{x}+\hat{\mathbf{j}},
			i}\, U^\dagger_{\mathbf{x},j}, \dots \right)\,,
	\end{split}
\end{equation}
to which we can apply the original derivative and obtain
\begin{align}
	\im \frac{\partial}{\partial \varphi_{\mathbf{x},1}} \psi & = \im
	\left(
	\frac{\partial}{\partial \widetilde{\varphi}_{\mathbf{x},1}} +
	\frac{\partial}{\partial
		\widetilde{\omega}_\mathbf{x}} -
	\frac{\partial}{\partial
		\widetilde{\omega}_{\mathbf{x} - \hat{\mathbf{2}}}} \right)
	\widetilde{\psi}
	\intertext{as well as}
	\im \frac{\partial}{\partial \varphi_{\mathbf{x},2}} \psi & = \im
	\left(
	\frac{\partial}{\partial \widetilde{\varphi}_{\mathbf{x},2}} -
	\frac{\partial}{\partial
		\widetilde{\omega}_\mathbf{x}} +
	\frac{\partial}{\partial
		\widetilde{\omega}_{\mathbf{x} - \hat{\mathbf{1}}}} \right)
	\widetilde{\psi} \, .
\end{align}
Note that the two electric operators differ in the signs in front of the
plaquette derivatives. As a consequence, the newly
introduced plaquette degrees of freedom do not appear in Gauss' law. It reads
\begin{equation}
	\im \left( \frac{\partial}{\partial \widetilde{\varphi}_{\mathbf{x},1}}
	+ \frac{\partial}{\partial \widetilde{\varphi}_{\mathbf{x},2}} -
	\frac{\partial}{\partial\widetilde{\varphi}_{\mathbf{x} -
			\hat{\mathbf{1}},1}}
	- \frac{\partial}{\partial \widetilde{\varphi}_{\mathbf{x} -
			\hat{\mathbf{2}},2}} \right) \widetilde{\psi} = 0 \, .
		\label{eq:gauss_law_dynamic_u1}
\end{equation}
Applied to the momentum basis we thus get
\begin{equation}
	m_{\mathbf{x},1} + m_{\mathbf{x},2} - m_{\mathbf{x} -
			\hat{\mathbf{1}},1} - m_{\mathbf{x} -
			\hat{\mathbf{2}},2} = 0 \,.
			\label{eq:gauss_law_momentum_u1}
\end{equation}
Therefore, it is left unchanged from the original Kogut-Susskind formulation.

Overall we have $N_{\textrm{sites}}
+ N_{\textrm{plaquettes}} = 2 \ell (\ell-1) + 1$ linear constraints acting on the $2\ell(\ell-1)$ helper links, in the open boundary case.
Both, the plaquette constraints~\eqref{eq:plaquette_constraints} and the Gauss' law constraints~\eqref{eq:gauss_law_momentum_u1} share the null-space of all $m_{\mathbf{x},i}$ equal.
All the remaining $2 \ell (\ell-1) - 1$ constraints are linearly independent.
Thus, the only solution is given by $m_{\mathbf{x}, i} = \text{const.}$ for all helper links.
Global gauge invariance allows us to choose all $m_{\mathbf{x}, i} = 0$ without loss of generality.
I.e.\ all helper links are forced into the electric ground state and will
not contribute to the Hamiltonian. We end up with the well known U(1) rotor
Hamiltonian from \cref{eq:u1dual}:
\begin{equation}
	\hat{H}_{\textrm{dual}} = \frac{g^2}{2} \sum_{\mathbf{x},i} \left(
	\frac{\partial}{\partial \omega_{\mathbf{x}}} -
	\frac{\partial}{\partial
		\omega_{\mathbf{x}+\hat{\mathbf{i}}}} \right)^2 +
	\frac{2}{g^2} \sum_{\mathbf{x}} \left( 1 -
	\cos\omega_\mathbf{x} \right) \,.
	\label{eq:u1_dual_dynamic}
\end{equation}
As this Hamiltonian is already well established we can be confident it will
yield the correct physics. In the following we will refer to this approach as a
\emph{plaquette separation} ansatz.

In the open boundary case, this does not offer anything new. However,
when considering periodic boundary conditions or the introduction of matter
fields, this approach remains local. This comes at the cost of having
to simulate in a much larger Hilbert space. In such simulations one would
ideally use a hybrid digitisation scheme. The plaquette dof would be
expanded in the plaquette states, while the helper links are
efficiently expressed in the electric basis. It remains to be seen, whether
this approach offers any advantage over non-local simulations in the rotor
approach.

\subsection{Attempts at a Dynamic Constraint in SU$(N)$}

Next we want to study the plaquette derivative constraint from
\cref{eq:u1-plaq-constraint} for a non-Abelian gauge group.

To begin with, we will consider a simplified setup: A single plaquette
of two helper links in an SU$(N)$ lattice gauge theory:
\begin{equation}
	P = U_1 U_2 \, .
\end{equation}
Again we want the helper link wave function to be independent of $P$. This will
be enforced by constructing a plaquette Lie derivative in the following way:
Similarly to measuring the change in the wavefunction caused by the
infinitesimal rotation of an individual link, one can also measure the change
caused by an infinitesimal rotation of the plaquette. Such a Lie derivative
takes the form of
\begin{equation}
	\hat{L}^c_{P} \, \psi = - \im \frac{\md}{\md \alpha} \,
	\psi(\mathrm{e}^{-\im \alpha \tau_c} \widetilde{U}_1, \,
	\widetilde{U}_1^\dag \mathrm{e}^{-\im \alpha \tau_c} \widetilde{U}_1
	\widetilde{U}_2) \Big|_{\alpha=0}
	\label{eq:Lp_non-abelian}
\end{equation}
Note, that for the second link the rotation  matrix $\mathrm{e}^{-\im \alpha
		\tau_c}$ first needs to be parallel transported along
$\widetilde{U}_1$, in
order to rotate plaquette correctly.

To express this Lie derivative in terms of the existing momentum and link
operators we can further simplify this expression:
\begin{align}
	\begin{split}
		\hat{L}^c_{P} \, \psi  & = - \im \frac{\md}{\md \alpha} \,
		\psi(\mathrm{e}^{-\im \alpha \tau_c}
		\widetilde{U}_1, \widetilde{U}_2) \Big|_{\alpha=0} \\
		& \qquad - \im \frac{\md}{\md \alpha} \, \psi(\widetilde{U}_1,
		\, \widetilde{U}_1^\dag
		\mathrm{e}^{-\im \alpha \tau_c} \widetilde{U}_1
		\widetilde{U}_2) \Big|_{\alpha=0}
	\end{split}
	\\
	 & = \left( \hat{L}^c_1 + \mathcal{R}^c_{\,c'}(\widetilde{U}_1)
	\hat{L}^{c'}_2
	\right) \psi
\end{align}
where
\begin{equation}
	\mathcal{R}^c_{\,c'}(U) = 2 \mathrm{Tr} \left[ \tau_c U \tau_{c'}
		U^\dag \right] \, .
\end{equation}
This operator also fulfils the canonical commutation relations (up to
a constant factor)
\begin{align}
	\left[ \hat{L}^c_{P} , P \right]\,               & = - 2 \tau_c P \,,
	\\
	-\mathcal{R} (P^\dag)^c_{\, c'} \hat{L}^{c'}_{P} & = \hat{R}^c_{P}
	\intertext{and}
	\left[ \hat{R}^c_{P} , P \right]\,               & = 2 P \tau_c \, .
\end{align}
The factor 2 appears because the plaquette here is the product of two links. A
plaquette of $n$ links will pick up a factor $n$.

The constraint removing the plaquette dof from the helper links then
reads
\begin{equation}
	\left( \hat{L}^c_1 + \mathcal{R}^c_{\,c'}(\widetilde{U}_1)
	\hat{L}^{c'}_2
	\right) \ket{\psi} = 0\,,\quad\forall c \, .
\end{equation}
By multiplying with $\mathcal{R}(\widetilde{U}_1^\dag)$ it simplifies to:
\begin{equation}
  \label{eq:Cplaq}
  \hat{C}^c_{\textrm{plaq}} \ket{\psi} =
  \left(-\hat{R}^c_1 + \hat{L}^c_2  \right) \ket{\psi} = 0\,,\quad\forall
  c\, .
\end{equation}
The physical interpretation of this constraint matches the constraint in U(1).
Again we sum over the charges of each link in our plaquette. The only
difference here, is that we need to parallel transport all the measured charges
to a mutual vertex. This is due to the non-Abelian nature of the gauge group.
Note that the minus sign in front of $\hat{R}^c_1$ is due to the fact
that the plaquette is a directed object, and, thus, unavoidable.

With these prerequisites, we can prove the following
\begin{theorem}[Inconsistency of dynamic plaquette constraint]
  The plaquette constraint on the helper links defined in
  \cref{eq:Cplaq} can be only fulfilled by states
  \begin{equation}
    \psi(\widetilde{U}_1, \widetilde{U}_2) = \mathrm{const}\, .
  \end{equation}
  Therefore, the helper link degrees of freedom are completely removed,
  and $\hat{C}^c_{\textrm{plaq}}$ overconstrains the system.
\end{theorem}

\begin{proof}
  Let $\ket{\psi}$ be a state fulfilling \cref{eq:Cplaq}.
  Consider the commutator
  \begin{equation}
    \label{eq:CComm}
    \left[ \hat{C}^a_{\textrm{plaq}} ,  \hat{C}^b_{\textrm{plaq}} \right] =
    \im f_{abc} \left( \hat{R}^c_1 + \hat{L}^c_2	\right)
  \end{equation}
  with the structure constant $f_{abc}$ as in \cref{eq:L,eq:R}, in which
  the minus sign from the plaquette orientation disappears.
  The commutator, on the other hand, applied to the state
  $|\psi\rangle$ gives
  \begin{equation}
  \left[ \hat{C}^a_{\textrm{plaq}} ,                                                 
    \hat{C}^b_{\textrm{plaq}} \right]|\psi\rangle = 0\,.
  \end{equation}
  Combined with \cref{eq:Cplaq} one obtains
  \begin{equation}
    \left(	- \hat{R}^c_1 + \hat{L}^c_2  \right)
    \ket{\psi} = \left(
    \hat{R}^c_1 + \hat{L}^c_2  \right) \ket{\psi} = 0\,,
  \end{equation}
  and, therefore
  \begin{equation}
    \hat{R}^c_1 \ket{\psi} = \hat{L}^c_2
    \ket{\psi} = 0 \qquad\forall\quad c\, .
  \end{equation}
  Thus,
  \begin{equation}
    \psi(\widetilde{U}_1, \widetilde{U}_2) = \mathrm{const}\,,
  \end{equation}
  as claimed.
\end{proof}
It should be clear that these constraints can, therefore, not lead to
physical results. However, it is still instructive to illustrate it
with the following simplified system. Consider the system of gauge links
depicted below:
\begin{center}
  \includegraphics[width=.3\columnwidth]{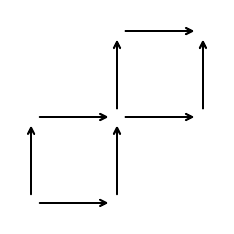}
\end{center}
Using Gauss' law at all sites, but the central one, we can \emph{glue} gauge
links together and reduce the system to only two gauge degrees of freedom:
\begin{center}
  \includegraphics[width=.3\columnwidth]{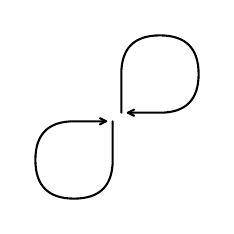}
\end{center}
The Kogut-Susskind Hamiltonian then reads:
\begin{equation}
  \hat{H} = 2 g^2 \left( |\hat{\vec{L}}_{1}|^2 + |\hat{\vec{L}}_{2}|^2
  \right) + \frac{2}{g^2} \left(4 - \mathrm{Tr} \left[ \hat{U}_1
    \right]  - \mathrm{Tr} \left[ \hat{U}_2 \right]\right) \, .
\end{equation}
It is simply the sum of two single plaquette Hamiltonians with no interactions
between the two gauge degrees of freedom $\widetilde{U}_1$ and
$\widetilde{U}_2$. Therefore, the eigenstates of this
Hamiltonian must be products of eigenstates of the single plaquette
Hamiltonian.
However, Gauss' law here differs from the single plaquette system:
\begin{equation}
  \left(\hat{L}^c_1 + \hat{R}^c_1 + \hat{L}^c_2 + \hat{R}^c_2 \right)
  \ket{\psi} = 0 \,.
\end{equation}
Thus also products of states which violate the single plaquette Gauss' law show
up in the physical spectrum of this system. These states are not possible in
the plaquette separated version of the system:
\begin{center}
  \includegraphics[width=.3\columnwidth]{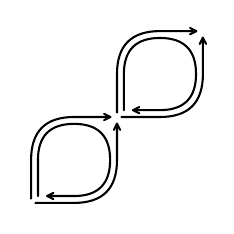}
\end{center}
When all helper links are forced into the electric ground state, we simply end
up with the single plaquette Gauss' law constraint for both
plaquettes.

For plaquettes consisting of more than two links one can repeat
the arguments with all but two of the links set to identities. 
Thus, a construction inspired by the U$(1)$ case fails in general.
We conclude that the plaquette separation Ansatz does not
work in SU$(N)$.

\section{Discussion}

First, let us discuss the practical consequences of the somewhat abstract
results presented in 
\Cref{th:non-locality} and \cref{th:non-locality-pbc}: if the simulation, on a quantum device or with
tensor network methods, is performed based on the original lattice, it
might not seem too relevant that the maximal distance on the tree
grows linearly in $\ell$. But this is not true: a generic $2$-point
$1$-local term $\hat O_{\mathbf{x}}\, \hat O_{\mathbf{x}+\hat i}$ in the
original Hamiltonian for instance will be translated to the maximal
tree as a string of operators
\[
\mathcal{S}_n(\mathbf{x}, \mathbf{x}+\hat i) = \hat{O}'_{\mathbf{x}}\cdot\hat{O}'_{\mathbf{y_0}}\cdots
\hat{O}'_{\mathbf{y}_{n-1}}\cdot \hat{O}'_{\mathbf{x}+\hat i} \,.
\]
Here, $\mathbf{y}_k\neq\mathbf{x}$
and $\mathbf{y}_k\neq\mathbf{x}+\hat i$ for $0\leq k < n$, and the
$\mathbf{y}_k$ prescribe a loop-free path on the tree from
$\mathbf{x}$ to $\mathbf{x}+\hat i$.
\Cref{th:non-locality} implies that there is at least one such
path with the tree distance $d_T(\mathbf{x}, \mathbf{x}+\hat i)
= \Omega(\ell)$, which also implies $n=\Omega(\ell)$.
That is, the number of operators in a single term grows at least linearly in the lattice extent.
Such a string
must be present, because otherwise the sites $\mathbf{x}$ and
$\mathbf{x}+\hat i$ would not interact any more, see also
Ref.~\cite{PhysRevD.107.074504} and our discussion in
\cref{sec:theory}.

Hence, such terms will likely increase the scaling of tensor network
simulations in an unacceptable way. For quantum simulations they will
dramatically increase the circuit depth of relevant quantum circuits,
while still not hampering a possible exponential improvement~\cite{Jordan:2012xnu,Hariprakash:2023tla,Rhodes:2024zbr,Draper:2026bcj,Ostmeyer:2026ckx}
of quantum computers over classical computers for the (time evolution)
simulation of lattice gauge theories.

Recall that we did not encounter any non-localities in the case of U(1), at least in 2D with open boundary conditions, see eqs.~\eqref{eq:u1dual} and~\eqref{eq:u1_dual_dynamic}.
The Abelian nature of the Lie group allowed us to write Gauss' law in the simple way~\eqref{eq:gauss_law_dynamic_u1} and reduce it to a linear system of equations~\eqref{eq:gauss_law_momentum_u1}.
Due to the linearity, the problem is exactly solvable and all the non-local operator strings can be eliminated.
Intuitively speaking, in Abelian theories the intermediate operators $\hat{O}'_{\mathbf{y_k}}$ play less of a role because we can commute $\hat{O}'_{\mathbf{x}+\hat i}$ all the way to be adjacent to $\hat{O}'_{\mathbf{x}}$.
Now, in the non-Abelian case factors like $\widetilde{U}_1^\dag \mathrm{e}^{-\im \alpha \tau_c} \widetilde{U}_1$ in equation~\eqref{eq:Lp_non-abelian} introduce a non-linearity making general solutions unlikely if not impossible.
Thus, we expect that the (at least) linearly non-local operator strings cannot be eliminated in non-Abelian gauge theories.

When it comes to dynamic constraints instead, it is difficult to make general
statements. However, we can show a
fundamental inconsistency in non-Abelian gauge theories for the equivalent to the
dynamic constraint leading to correct results in U(1).
The inconsistency is
mainly due to \cref{eq:CComm}: the constraints
$\hat{C}_\mathrm{plaq}^a$ do not form an algebra. From this result one
might conjecture that a valid constraint must form an algebra,
however, we have no indication that this is possible. 

As a consequence of the discussion above, we conclude that the
local Hamiltonian derived in Ref.~\cite{PhysRevD.107.074504} is not
implementable in practice without introducing equivalent
non-localities, if the constraint this Hamiltonian is subject to
is supposed to be static in the sense of this paper. Any dynamic
implementation of this constraint beyond the one we have shown to be
inconsistent might be possible, but remains unknown.

\section{Outlook}

An approach to lattice gauge theories that avoids the problems in non-Abelian
lattice gauge theories discussed in this paper altogether is the one via
quantum link
models~\cite{Horn:1981kk,Orland:1989st,Chandrasekharan:1996ih,Brower:1997ha}.
Simulations are performed with an additional, unphysical dimension added and
the target gauge theory is obtained via dimensional
reduction. However, the mechanism of dimensional reduction has been
tested so far only in CP(N)
models~\cite{Beard:2004jr,Beard:2006mq,Laflamme:2015wma,Evans:2018njs},
and a practical test in U(1) or SU(N) gauge theories must be
investigated in the future.
Still, interest in quantum link models is growing: for U(1) recent
work can be found for instance in
Refs.~\cite{Dey:2026nxk,Gandon:2026das,Dey:2025eyq}.
A first
investigation of strings in a non-Abelian SU(2) quantum link model in the
Hamiltonian formulation using tensor network methods can be found in
Ref.~\cite{Ludwig:2026gqf}, and a Monte Carlo study of a similar model
in Ref.~\cite{Banerjee:2017tjn}.

Neural network wavefunctions have also attracted attention recently,
see for instance Ref.~\cite{Spriggs:2025sea}. With those, the problems
described here can be circumvented. However, for results
beyond ground states more work is required.

Beyond that, the authors are currently not aware of any method that allows for
efficient and local simulations at weak couplings. We believe that the class of
possible dynamic constraints deserves some further
exploration. However, if no valid candidate
can be found, long range interaction terms will likely need to be accepted and
dealt with for weak coupling Hamiltonian simulations of non-Abelian lattice
gauge theories.

\begin{acknowledgments}
  We thank Karl Jansen and Simone Romiti for many helpful discussions,
  comments, and the most enjoyable collaboration. We thank Petra
  Mutzel for communication on a draft of this manuscript.
  This project was funded by the Deutsche Forschungsgemeinschaft (DFG,
  German Research Foundation) as a project in the CRC 1639 NuMeriQS --
  project no.\ 511713970, and under Germany’s Excellence Strategy –
  EXC 3107 – Project-ID 533766364 in the Color-meets-flavor cluster of
  excellence.
We thank ECT* for support at the Workshop \enquote{Hamiltonian Lattice Gauge
    Theories: Status, Novel Developments and Applications} during which this work
  has been developed.
We have used ChatGPT to check the manuscript for typographical and
  logical mistakes, and Gemini to propose relevant references from the
  field of graph theory.
\end{acknowledgments}

\appendix

\section{Additional proofs}\label{sec:proofs}

\subsection{Continuation of the proof of \cref{th:max-tree}}\label{sec:proof_max-tree}

\begin{proof}	
	Part (c) periodic: For $d\ge 3$, let us first consider periodic boundary conditions. Every lattice site has $d$ ``positive'' directions. Thus, if there are $N$ lattice sites, then there are $Nd$ many links. However, each pair of outgoing links defines a plaquette, i.e., there are ${d\choose 2}=\frac{d(d-1)}{2}$ many plaquettes. For $d=3$ this means that there are exactly the same number of links as plaquettes and for $d\ge4$ there are more plaquettes than links. Hence,  there are at least as many rotor degrees of freedom as link degrees of freedom. Attempting to remove as many link variables as rotor variables can be introduced therefore either leaves no remaining helper links ($d=3$) or cannot be achieved at all ($d\ge4$). Having no remaining helper links necessarily disconnects the lattice, thus changing the underlying physics again.
	
	Part (c) open: In the case of open boundary conditions, let us first consider the smallest such hyper-cubic lattice, the $d$-dimensional cube. The $d$-dimensional cube has $2^d$ many vertices $\{0,1\}^d$. Each edge corresponds to a vertex $v=(v_1,\ldots,v_d)\in\{0,1\}^d$ and picking a coordinate $j$ flipping $v_j\mapsto 1-v_j$. Thus, there are $d2^d$ many edge generating operations and, since the edges are undirected, there are $d2^{d-1}$ links in the smallest hyper-cubic lattice. Similarly, we obtain the number of $k$-faces by picking $k$ different coordinate flips and accounting for the $2^k$-fold symmetry, i.e., there are ${d\choose k}2^{d-k}$ many $k$-faces in the $d$-dimensional hypercube. Since the plaquettes are precisely the $2$-faces, we obtain that there are $p(d)=d(d-1)2^{d-3}$ many plaquettes. The connectivity requirement implies that at least $2^d-1$ many helper links must remain, i.e., we can at most introduce $r(d)=(d-2)2^{d-1}+1$ many rotor variables. The function $p(d)-r(d)$ is $1$ at $d=3$ (there are $6$ plaquettes but only $12-7=5$ removable links) and strictly increasing for $d\ge3$. Hence, there are no $d$-dimensional hypercubes in which there are sufficiently many removable links to introduce all rotor degrees of freedom.
	
	Finally, we can complete the proof by showing that the number of plaquettes grows faster than the number of links. Considering the $d=3$ case, let us assume we have an $\ell_1\times\ell_2\times \ell_3$-size lattice and we want to enlarge it to be an $\ell_1\times\ell_2\times (\ell_3+1)$-size lattice. This enlargening introduces $\ell_1\ell_2$ many links in the third direction and an additional $\ell_1\times\ell_2$-size constant-$\ell_3$ slice. From part (a) we know that the additional links and plaquettes in the constant-$\ell_3$ slice balance exactly, so we only need to consider the new plaquettes containing a new link in the third direction. There are $(\ell_1-1)\ell_2$ many plaquettes closing in the first direction, and $\ell_1(\ell_2-1)$ many closing in the second direction. Hence, we have $\ell_1\ell_2$ many new links but $2\ell_1\ell_2-\ell_1-\ell_2$ many new plaquettes. Hence, there are $\ell_1\ell_2-\ell_1-\ell_2=(\ell_1-1)(\ell_2-1)-1\ge0$ many additional plaquettes. The initial shortfall of the $2\times2\times2$-lattice cannot be overcome and by induction all $d=3$ hyper-cubic lattices have too many rotor degrees of freedom.
	
	For $d\ge4$ we can perform the same analysis using an $\ell_1\times\cdots\times\ell_d$ lattice that extends to an $\ell_1\times\cdots\times\ell_{d-1}\times(\ell_d+1)$ lattice. Again, we obtain $\prod_{j=1}^{d-1}\ell_j$ many links in the final direction and a copy of the constant-$\ell_d$ slice. The constant-$\ell_d$ slice is a $(d-1)$-dimensional hyper-cubic lattice which by induction as $d-1\ge3$ has more plaquettes than available link variables. Furthermore there are $(\ell_j-1)\prod_{k=1,k\ne j}^{d-1}\ell_k$ many plaquettes in the $(j,d)$ direction. Thus, there are $\prod_{k=1}^{d-1}\ell_k$ many new links in the extension direction but $\sum_{j=1}^{d-1}(\ell_j-1)\prod_{k=1,k\ne j}^{d-1}\ell_k$ many new plaquettes. In other words, there are at least $(d-2)\prod_{k=1}^{d-1}\ell_k-\sum_{j=1}^{d-1}\prod_{k=1,k\ne j}^{d-1}\ell_k$ many additional rotor degrees of freedom more than additional link degrees of freedom. Setting $L:=\prod_{k=1}^{d-1}\ell_k$ we can simplify the expression of excess rotor variables as $(d-2)L-\sum_{j=1}^{d-1}\frac{L}{\ell_j}=L\left(d-2-\sum_{j=1}^{d-1}\ell_j^{-1}\right)$ and since open boundary conditions imply $\ell_j\ge2$, this yields that the excess rotor degrees of freedom are at least $L\frac{d-3}{2}>0$. Induction over all lattice sizes at fixed $d$ followed by an induction over $d$ implies that there are always more rotor degrees of freedom than link degrees of freedom for $d\ge4$.
\end{proof}
 
\end{document}